\documentclass[journal,10pt]{IEEEtran}

\usepackage{cite}
\usepackage{amsmath,amssymb,amsfonts}
\usepackage{algorithmic}
\usepackage{graphicx}
\usepackage{textcomp}
\usepackage{graphicx}                                            
\usepackage{multicol}         
\usepackage[font=small]{caption}
\usepackage{amssymb,amsmath,dsfont}
\usepackage{inputenc}
\usepackage{bbm}
\usepackage{cite}
\usepackage{color}
\usepackage{multirow}
\usepackage{subcaption}
\usepackage{array}
\usepackage{algorithm}
\usepackage{color,soul}
\usepackage{blindtext}
\usepackage{bm}
\usepackage[sort&compress,numbers]{natbib}
\usepackage{mdframed}
\setcitestyle{square}

\makeatletter
\newcommand{\mathleft}{\@fleqntrue\@mathmargin0pt}
\newcommand{\mathcenter}{\@fleqnfalse}
\makeatother

\newtheorem{proof}{Proof}
\newtheorem{remark}{Remark}

\begin{document}

\title{Joint data rate and EMF exposure analysis in Manhattan environments: stochastic geometry and ray tracing approaches}

\author{Charles~Wiame,~\IEEEmembership{Member,~IEEE}, Simon~Demey,~\IEEEmembership{Member,~IEEE}, Luc~Vandendorpe,~\IEEEmembership{Fellow,~IEEE}, Philippe~De Doncker,~\IEEEmembership{Member,~IEEE}, and~Claude~Oestges,~\IEEEmembership{Fellow,~IEEE}

\thanks{This work has been submitted to the IEEE for possible publication. Copyright may be transferred without notice, after which this version may no longer be accessible.}
\thanks{Manuscript received XXX, XX, 2022; revised XXX, XX, 2022.}}

\markboth{IEEE Transactions on Vehicular Technology,~Vol.~XX, No.~XX, XXX~2022}
{}

\maketitle

\begin{abstract}
The objective of this study is to jointly analyze the data rate and electromagnetic field (EMF) exposure in urban environments. Capitalizing on stochastic geometry (SG), a network level analysis is performed by modelling these environments via Manhattan Poisson line processes (MPLP). Using this framework, a number of performance metrics are derived: first moments, marginal distributions and joint distributions of the data rate and exposure. In addition, the original Manhattan model is generalized to include advanced features: corner diffraction, presence of potential blockages in streets, and users positioned at crossroads. As a second approach, deterministic ray tracing (RT) is utilized to compute the same metrics. The two methods are shown to provide close results on the condition that the model parameters are coherently selected. Furthermore, the numerical results enable to gain insight into several aspects: the role of the propagation mechanisms in the performance metrics, existing trade-offs between the rate and exposure requirements, as well as the impact of the user location (at a crossroad or in a single street).
\end{abstract}

\begin{IEEEkeywords}
EMF exposure, stochastic geometry, ray tracing, Manhattan Poisson line process.
\end{IEEEkeywords}

\IEEEpeerreviewmaketitle

\section{Introduction}
From Bell Labs perspective, the global data traffic is expected to skyrocket in future wireless networks \cite{weldon2016future}. The increasing number of mobile devices and the growing usage of data-consuming applications are two of the main factors supporting this prediction \cite{ali2018modeling}. In order to satisfy the resulting increase in capacity demand, \cite{claussen} identifies the densification of base stations (BS) as a promising solution. One possible manner to implement this densification consists in deploying small cell BSs to offload macro BSs. This approach is shown to achieve better spectral efficiency and to extend coverage regions \cite{osseiran20165g}. According to the study in \cite{ge20165g}, it is expected to find up to 40-50 BSs/km² to meet the coverage requirements expected for 5G and beyond. However, such intensive densification may raise questions in terms of sustainability, cost, and human exposure (for which legal thresholds must be respected in some countries). The analysis of this study focuses on this last exposure aspect. \\

The first areas where the densification is likely to be planned are city centers, which feature a particularly high user activity. Compared to other environments (suburbs or rural areas), these sites are also characterized by an important density of blockages impacting the propagations conditions. Channels models characterizing these environments should hence include this blockage aspect with accuracy. A number of techniques have been investigated to this purpose: ray-tracing (RT) \cite{geok2018comprehensive}, Winner-type models \cite{bultitude20074}, COST-type models \cite{damosso1999cost}, etc. Each of these approaches has its own specificities and is more or less suitable for a network-level analysis. On the one hand, some of the deterministic tools (e.g. ray tracing) have the advantage of modelling the physical propagation with accuracy, but they often require important computational resources \cite{RTtuto}. On the other hand, BS densification is likely to induce additional randomness in future networks: in the user activity, node positions, etc. This justifies the use of stochastic models, which may be of interest at a large network scale. From a global network perspective, some of the link-level details are often abstracted or simplified when using such methods. This enables to meet computational constraints, but comes at the cost of simplifications in the propagation mechanisms. Stochastic geometry (SG) is one of the tools enabling to study the global performance at such network level \cite{elsawy2016modeling}. This branch of spatial statistics abstracts base stations and users as random point processes. On the basis of these processes, analytical expressions of metrics can be derived: coverage probability, spectral efficiency, etc. These expressions are averaged out over all the potential node locations, and therefore capture the full spatial statistics. Such an approach therefore enables to circumvent brute force Monte Carlo (MC) simulations. Within the SG theory, two point processes are employed to model urban topologies: the Manhattan Poisson line process \cite{Baccelli_urban} (MPLP, used to represent streets with a regular grid structure), and the Cox process \cite{choi2018poisson} (use to represent cities with irregular street deployments). This work considers the former topology, and attempts to compare results obtained via RT and SG for this urban environment.


\subsection{Related works}
The study in \cite{Baccelli_urban} is one of the first SG works employing the MPLP to model roads. The coverage distribution is derived using a correlated blockage model accounting for building penetration. In \cite{wang2018mmwave}, the same metric is analyzed for vehicle-to-infrastructure networks operating at mmWave frequencies. The MPLP is shown to accurately describe Manhattan networks by comparison with the fixed grid model and the actual street deployment in Chicago. MPLP models are also used for indoor scenarios \cite{akin2018effect,zhang2015indoor,muller2016analyzing}, drones-to-ground networks \cite{hriba2021optimization} and relay assisted networks \cite{elbal2019coverage}. Other works also consider Cox line processes as generalization of the MPLP to model non-prependicular streets \cite{chetlur2020load,chetlur2018coverage,choi2018poisson,choi2021modeling,chetlur2019coverage,chetlur2018success,choi2018analytical,ghatak2019small}. Among the different technologies that have been studied, one can mention vehicle-to-everything communication \cite{chetlur2020load}, vehicular ad hoc networks \cite{chetlur2018success} and multi-radio access technology \cite{ghatak2019small}. In \cite{chetlur2020load}, trends on the network deployment design have been highlighted and the benefit of small cells to offload macro BSs has been proven.     \\
  
Many works focus on the analysis of emerging networks using pure SG approaches. However, a very limited number of publications propose validations of these SG models using other methods (comparison to datasets, measurements, or softwares capturing additional electromagnetic phenomenons such as diffraction or scattering, etc). In \cite{lu2015stochastic}, an experimental validation is conducted using two open access databases of mobile operators in United Kingdom. More recently, \cite{ram2021optimization} validated a radar detection analysis of using finite difference time domain technique. To the best of the authors knowledge, no validation of SG models using other software tools (e.g. RT) can be found in the literature. 

\subsection{Contributions}

On the basis of the aforementioned works, the main findings of this study can be summarized as follows: 

\begin{itemize}

\item The Manhattan model originally presented in \cite{Baccelli_urban} is fully revisited. A number of features are added to this model to capture additional propagation aspects. These new characteristics are detailed below:

\begin{itemize}
    \item The proposed analysis takes into account power contributions coming from BSs outside the user street and propagating via diffraction at crossroad corners. As explained in \cite{wang2018mmwave}, contributions from such a corner diffraction can be significant compared to building penetration, hence the need to address their modelling. To this purpose, Berg recursive method is employed for its simplicity and its accuracy regarding the path loss (PL) levels \cite{berg1995recursive}. This model also has the advantage of featuring a general expression, depending on both Euclidean distances and corner angles. These dependencies open perspectives for a generalization to Cox processes, and differ from \cite{wang2018mmwave} where constant diffraction losses are considered. 
    \item Unlike \cite{Baccelli_urban} where the small scale fading was only Rayleigh, our framework enables to incorporate any fading distribution in the model. 
    \item The presence of street obstacles (cars, trucks,…) is included using the blockage model of \cite{jarvelainen2016evaluation}, and integrated in the SG framework via inhomogeneous thinning. This thinning is function of a line-of-sight (LOS) probability which is distance dependent and continuous everywhere, unlike the LOS ball model \cite{singh2015tractable}.
    
    \item In practical deployments, a non-negligible proportion of users can be located at street intersections (e.g. pedestrians waiting at traffic lights). In such cases, the simultaneous presence of BSs in the two crossing streets results in power levels differing from those measured at other users. In the original work of \cite{Baccelli_urban}, users were almost surely never located at street intersections. Conversely, our work includes a non-zero probability for this event to happen, in which case power signals coming from the two perpendicular streets are considered.  
    
    \item The heights of the UEs and BSs are here considered. Neglecting this feature might result in underestimations in the PLs, especially for BSs in the user vincinity. The power contributions associated to these BSs might hence be overestimated, which could bias the performance metrics.
\end{itemize}

\item To the best authors' knowledge, this analysis is the first work studying EMF exposure in Poisson line processes. In order to characterize the statistics of exposure and its correlation to the spectral efficiency, the following metrics are derived: the first moments of the data rate and exposure, their marginal distributions, as well as their joint distribution.

\item Most SG publications validate their analytical results using equivalent MC simulations. While such procedure enables to verify the correctness of the derived expressions, it does not provide indications on whether or not they are realistic compared to more physical propagation models. This study attempts to bridge the resulting gap by comparing our SG expressions to RT simulations reproducing the Manhattan environment. To our best knowledge, this paper is the first work performing such RT validation. From this comparison, we conclude that the two approaches can produce tight results. Selecting the propagation parameters of the SG framework appropriately is however a necessary condition to obtain an accurate matching. A method is proposed in section \ref{compRT} to properly tune these parameters. A sensitivity analysis is also performed in the same section to support these conclusions.

\end{itemize}
\subsection{Organization of the paper}
The system model and the analyzed metrics are defined in Section \ref{sect:system}. The SG expressions of these metrics are derived in Section \ref{anres}. The numerical results of the SG analysis are presented in Section \ref{numres}. A case study comparing the SG and RT frameworks is then detailed in section \ref{compRT}. Section \ref{ccl} finally summarizes the takeaways of this study, and provides future research directions.

\textit{Notations:} in the next sections, $j=\sqrt{-1}$ denotes the imaginary unit. $\operatorname{Im}\{\cdot\}$ represents the imaginary part of a complex number. $\mathbf{x}^H$ is the conjugate transpose of matrix $\mathbf{x}$.


\section{System Model}
\label{sect:system}

\subsection{Network topology}
One considers a street infrastructure deployed in a square area of length $2R$ and centered around $(0,0) \in \mathbb{R}^2$. This deployment is modelled by means of a MPLP: horizontal and vertical streets are built using two independent Poisson point processes (PPP) denoted by $\Phi_{VS}$ and $\Phi_{HS}$. The density of these PPPs is here constant and given by $\lambda_S$. The empty spaces between these streets are assumed to be buildings.

Base stations are generated in each street using independent PPPs of density $\lambda_B$. In the following sections, the total set containing all these BSs will be denoted as $\Phi_B$. These BSs operate at frequency $f$ and are placed at height $H_B$. In addition, they are all equipped with a single antenna of unit gain. One will denote as $P_B$ their constant and isotropic  transmit power. 

\begin{figure} []
    \centering
    \includegraphics[width=0.37\textwidth]{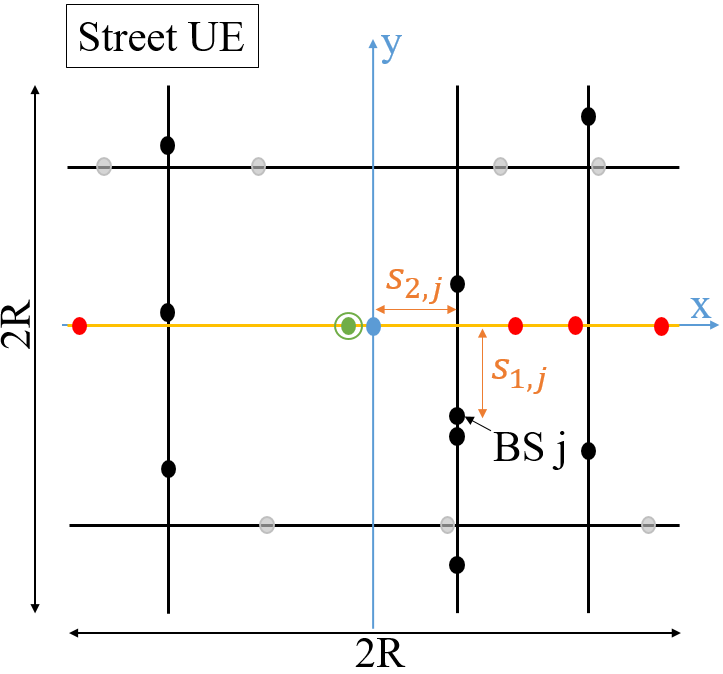}
    \par\vspace{0.5cm}
    \includegraphics[width=0.37\textwidth]{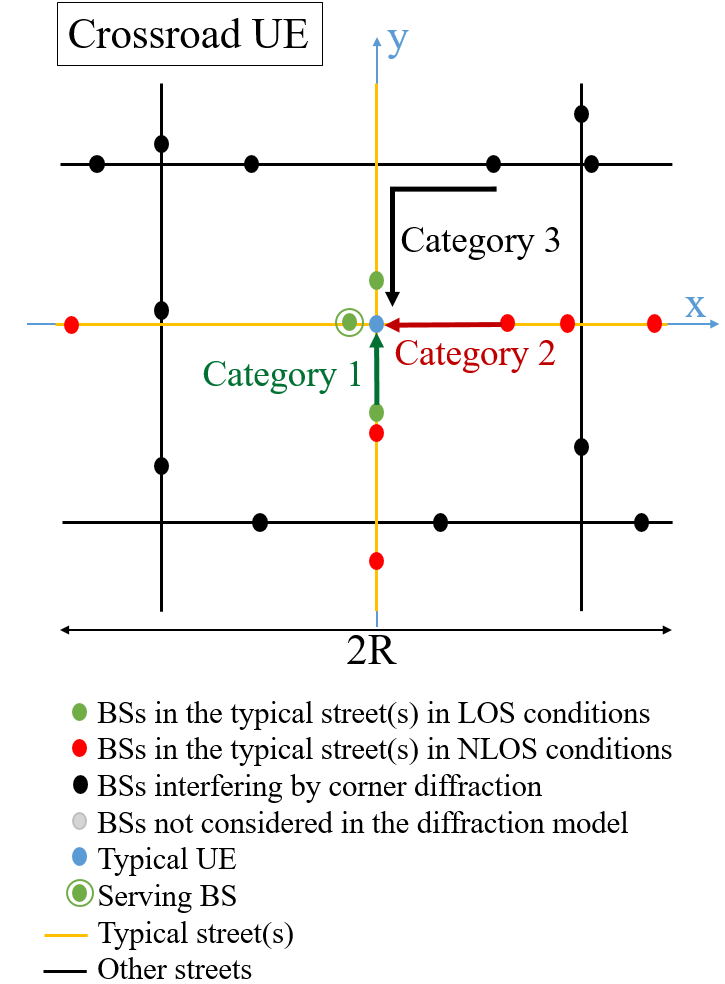}
    \par\vspace{0.5cm}
    \caption{Illustration of the topology in the cases of a street user and of a crossroad user.}
    \label{typesUEs}
\end{figure}

The network analysis is performed for a typical UE located at $(0,0, H_U )$. One will denote $\Delta H = H_B - H_U$, its height difference with respect to the BSs. In the framework of this work, this difference is assumed to be greater than 1 meter to avoid path loss singularities (see Subsection \ref{computation of received powers}). The UE is also equipped with a single isotropic antenna of unit gain. As discussed in the previous section, the location of this UE will strongly impact the received power levels: users located at crossroads are in average more exposed to LOS signal compared to users that are not located at street intersections. To take this feature into account, the analysis proposed in this study is performed in two steps:
\begin{enumerate}
    \item Street and crossroad UEs are first separately investigated. The performance metrics are independently derived for each of these two user types. In the case of the crossroad UE, two streets (of ordinate $y=0$ and abscissa $x=0$) are added to the MPLP to model the intersection (see Figure \ref{typesUEs}). By contrast, a single avenue (of ordinate $y=0$) is generated for the street UE (Figure \ref{typesUEs}). These additional lines in which the typical user is located will be referred to as \textit{typical streets} in the rest of this document.
    \item General metrics are deduced for an arbitrary UE possibly located at a crossroad with probability\footnote{This probability $\eta$ can for instance be defined as the ratio between the streets and building widths.} $\eta$ and in a street with probability $1-\eta$. The expressions for these final metrics are derived using a weighted average depending on this parameter $\eta$.
\end{enumerate}

\subsection{Association policy}

For sake of mathematical tractability, the typical user is associated to the closest BS of its typical street(s). The selected BS is hence the candidate in this typical street(s) minimizing the 1D distance to the user. The other BSs of $\Phi_B$ are assumed to provide interference. Under this policy, the distribution of the distance from a street user to its serving BS is known and provided in \cite{Wang}.
\begin{equation}
    \label{two_streets}
    f_{r,1}(r) = \dfrac{2\lambda_{B}e^{-2\lambda_{B}r}}{1 - e^{-2\lambda_{B}R}}, \; \; \text{for} \; \; 0 < r < R.
\end{equation}

For users located at crossroad, the distribution considers the presence of the two streets: 
\begin{equation}
    \label{two_streets}
    f_{r,2}(r) = \dfrac{4\lambda_{B}e^{-4\lambda_{B}r}}{1 - e^{-4\lambda_{B}R}}, \; \; \text{for} \; \; 0 < r < R.
\end{equation}


\begin{remark}
    A generalization to a strongest average power association is possible and left for future work. In the framework of this study, it is possible to show that the two rules lead to similar results for low blockage probabilities.
\end{remark}
\subsection{Blockage model}

Obstacles (vehicles, urban furniture, etc) are assumed to be present in the typical street(s). These blockages are assumed to induce non line-of-sight (NLOS) links for some of the BSs. In order to model this aspect, the probabilities used in \cite{jarvelainen2016evaluation} are employed: given a BS in the typical street(s) located at distance $r$ from the user, the probability for this BS to be in LOS is given by $p_{L}(r) = e^{-\beta r^\gamma}$, with $\beta,\gamma \in \mathbb{R}^+$. In a complementary way, the probability of NLOS is given by $p_{N}(r) = 1 - p_{L}(r)$.

\subsection{Computation of the received powers}
\label{computation of received powers}
The analysis of this study distinguishes three categories of received power signals. These categories are detailed hereafter and represented in Figure \ref{typesUEs}.

\subsubsection*{Category 1 - LOS BSs in the typical street(s)}

the power received at the typical UE $i^*$ from a LOS BS $j$ of the typical street(s) is given by
\begin{equation}
     P_{i^*j}^{(SG)} = P_B \left|h_{i^*j}^{(L)}\right|^2 \kappa_L^{-1} \sqrt{d_{i^*j}^2+\Delta H^2}^{-\alpha_L}.
     \label{PMC11}
\end{equation}
In this definition, $\left|h_{i^*j}^{(L)}\right|^2$ accounts for the small scale fading. $\kappa_L$ and $\alpha_L$ respectively represent the intercept and path loss exponent of the path loss model. $d_{i^*j}$ denotes one-dimensional distance from UE $i^*$ and BS $j$. 

\subsubsection*{Category 2 - NLOS BSs in the typical street(s)}
the expression of the powers received from these BSs is similar to category 1:
\begin{equation}
     P_{i^*j}^{(SG)} = P_B \left|h_{i^*j}^{(N)}\right|^2 \kappa_N^{-1} \sqrt{d_{i^*j}^2+\Delta H^2}^{-\alpha_N}
     \label{PMC12}
\end{equation}
The difference with the LOS category is related to the PL parameters $\kappa_N$ and $\alpha_N$, and the fading coefficients $\left|h_{i^*j}^{(N)}\right|^2$. These variables can take values/distributions different from $\kappa_L$, $\alpha_L$ and $\left|h_{i^*j}^{(L)}\right|$, enabling more severe attenuations on the NLOS links.

\subsubsection*{Category 3 - Corner diffraction}
the power contributions of this third category are associated to BSs located in avenues perpendicular to the typical street(s) (Fig. \ref{typesUEs}). These powers affect the typical UE via corner diffraction and are modelled using Berg recursive method \cite{berg1995recursive}. The power received from BS $j$ is in that case given by:  
\begin{equation}
      P_{i^*j}^{(SG)}  = P_B \left|h_{i^*j}^{(D)}\right|^2 \kappa_D^{-1} \mathcal{D}_b(s_{1,j},s_{2,j})^{-\alpha_D}
      \label{PMC2}
\end{equation}
where $\kappa_D$, $\alpha_D$ and $\left|h_{i^*j}^{(D)}\right|^2$ are respectively the PL and the fading power coefficients associated to this category of signals. The variable $\mathcal{D}_b(s_{1,j},s_{2,j})$ represents the Berg distance. For a unique building edge diffraction, it is defined as 
\begin{equation}
    \mathcal{D}_b(s_{1,j},s_{2,j}) = s_{1,j} + s_{2,j} + k_f \bigg(\dfrac{\theta_{i,j}}{\pi/2}\bigg)^{\nu} s_{1,j} s_{2,j}.
    \label{virtual_dist}
\end{equation}
In this definition, $k_f = \sqrt{\frac{q_{\lambda} f}{c}}$ where $q_{\lambda}$ is a frequency dependent model parameter and $c$ is the speed of light. If the system is operated at a frequency between 2 and 6 GHz, $q_{\lambda}$ can be fixed at 0.031 for urban micro-cells \cite{metis}. $\theta_{i,j}$ is the angle between the street of BS $j$ and the street of user $i^*$. Its value is always equal to $\pi/2$ in the case of the MPLP. $\nu$ is an empirical parameter which has no influence in this study. Finally, the variables $s_{1,j}$ and $s_{2,j}$ denote the distances from BS $j$ to the street corner and from the street corner to user $i^*$ (cfr Figure \ref{typesUEs}). 
\begin{remark}
one will note that $\mathcal{D}_b(\cdot,\cdot)$ is greater than the Euclidean distance from $i^*$ to $j$, owing to the diffraction loss.
\end{remark}
\begin{remark}
contributions associated to building penetration are here neglected. As mentioned in \cite{wang2018mmwave}, this assumption is particularly true for mmWaves. For scenarios where this hypothesis may not be valid, the inclusion of these contributions as fourth category is left for future works. In addition, equations \eqref{PMC2} and \eqref{virtual_dist} constitute the first order Berg model: signals diffracted more than once are not considered here.
\end{remark}
\begin{remark}
an exclusion zone is considered in the street generation to prevent the path loss in \eqref{PMC2} to be greater than 1. This zone is defined as a segment $[-r_s;r_s]$ around the typical user. Streets potentially generated in this zone are ignored. Regarding the other path loss models of \eqref{PMC11} and \eqref{PMC12}, potential singularities are circumvented thanks to parameter $\Delta H>1$.
\end{remark}
\begin{remark}
in this mathematical framework, all the fading coefficients can take any distribution of known characteristic function. In the numerical results presented in section IV, one will however focus on the particular case of Rice fading for categories 1 and 2, and Rayleigh fading for category 3.
\end{remark}

\subsection{Performance metrics}
The metrics studied in this work are detailed below.

\begin{itemize}
    \item The average user capacity is defined as 
    \begin{align}
         &\mu_C = \mathbb{E} \Big[B\log_{2}\left(1+\text{SINR}_{i^*}\right) \Big] \; \; \text{where} \\
         &\text{SINR}_{i^*} = \frac{P_{i^*j^*}}{\underset{\substack{j \in \Phi_B\\j\neq j^*}}{\sum}P_{i^*j} + W} = \frac{S}{I_L +I_N +I_D +W}.
    \end{align}
    In the above expressions, $B$ is the system bandwidth. $P_{ij}$ represents the power transmitted from BS $i$ to UE $j$. $j^*$ is the index of the serving BS. The variables $S$, $I_L$ and $I_N$ represent the useful signal power and the interference associated to LOS and NLOS BSs of the typical street(s). $I_D$ denotes the interference associated to corner diffraction. $W$ finally represents the constant noise power. 
    \item The average exposure coming from all network BSs is defined as 
    \begin{equation}
        \mu_E = \mathbb{E} \bigg[\sum_{j \in \Phi_B} P_{i^*j}\bigg] = \mu_L + \mu_N + \mu_D,
    \end{equation}
    where $\mu_{L}$, $\mu_{N}$ and $\mu_{D}$ are respectively the mean total powers coming from the typical street(s) in LOS condition, from the typical street(s) in NLOS condition, and associated to diffraction. With a slight abuse of notation, $S$ is here included either in $\mu_L$ or in $\mu_N$.
    \item The coverage probability is defined as 
    \begin{align}
        P_c(\theta_c) &= \mathbb{P}\big[\text{SINR}_{i^*} > \theta_c\big].
    \end{align}
     \begin{remark}
         the distribution of the capacity is defined as $P_{cap}(\theta) = \mathbb{P}\big[B\log_{2}\left(1+\text{SINR}_{i^*}\right) > \theta\big]$ and can directly be deduced via the change of variable $P_{cap}(\theta) = P_{c} (2^{\frac{\theta}{B}}-1)$.
     \end{remark} 
    \item the cumulative distribution function (CDF) of the exposure is defined as 
    \begin{align}
        &P_e(\theta_e) = \mathbb{P}\bigg[\sum_{j \in \Phi_B} P_{i^*j}< \theta_e \bigg] \\
        &\text{where} \quad \sum_{j \in \Phi_B} P_{i^*j} = S + I_L + I_N + I_D.
    \end{align}
    \item The joint SIR-exposure distribution is defined as 
    \begin{equation}
        F(\theta_c,\theta_e) = \mathbb{P}\bigg[\text{SINR}_{i^*} > \theta_c, \sum_{j \in \Phi_B} P_{i^*j}< \theta_e\bigg].
        \label{Joinprob}
    \end{equation}
    This distribution represents the probability for a mobile user to satisfy coverage requirements (SINR greater than $\theta_c$) while experiencing an exposure below a safety threshold $\theta_e$.
    \begin{remark}
        this metric differs from the joint distribution of \cite{Paper_JCCDF}, where the exposure is required to be above a threshold for energy harvesting purposes. To the best of the authors knowledge, no work in the literature has considered \eqref{Joinprob} at the moment of writing this study.
    \end{remark}
    
\end{itemize}

\begin{figure*}
\normalsize
\setcounter{equation}{15}

\begin{align}
\phi_{L,1}(t|r) &= \exp\Bigg[2\lambda_{B}\int_{r}^{R} p_{L}(r') \bigg[ \phi_F\Big(P_B\kappa_L^{-1}\sqrt{r'^2 +\Delta H^2}^{-\alpha_L}t;K\Big) - 1\bigg] dr' \Bigg] \label{eqcharfunL} \\
\phi_{N,1}(t|r)&= \exp\Bigg[2\lambda_{B}\int_{r}^{R} p_{N}(r')\bigg[ \phi_F\Big(P_B\kappa_N^{-1}\sqrt{r'^2 +\Delta H^2}^{-\alpha_N}t;K\Big) - 1\bigg] dr' \Bigg] \label{eqcharfunN} \\
\phi_{D,1}(t) &= \exp\Bigg[-2\lambda_S \int_{r_s}^{R} 1 - \exp\bigg[ -2\lambda_{B} \int_{0}^{R}1-\phi_F\Big(P_B\kappa_D^{-1}\mathcal{D}_b(x,y)^{-\alpha_D}t;0\Big)dx\bigg]dy \Bigg] \label{eqcharfunD} 
\end{align}
\setcounter{equation}{19}
\begin{align}
P_{c,q}(\theta_c) = & \dfrac{1}{2} + \dfrac{1}{\pi} \int_{0}^{\infty} Im\bigg[\phi_{D,q}(-\theta_c t) \int_{0}^{R}p_{L}(r)\phi_{S}^{(L)}(t|r)\phi_{L,q}(-\theta_c t|r)\phi_{N,q}(-\theta_c t|r)f_{r,q}(r)dr \bigg] t^{-1}dt \nonumber \\ 
&+ \dfrac{1}{\pi} \int_{0}^{\infty} Im\bigg[\phi_{D,q}(-\theta_c t) \int_{0}^{R}p_{N}(r)\phi_{S}^{(N)}(t|r)\phi_{L,q}(-\theta_c t|r)\phi_{N,q}(-\theta_c t|r)f_{r,q}(r)dr \bigg] t^{-1}dt \label{eqCovprob} \\
P_{e,q}(\theta_e) = & \dfrac{1}{2} - \dfrac{1}{\pi} \int_{0}^{\infty} Im\bigg[e^{-jt\theta_e}\phi_{D,q}(t) \int_{0}^{R}p_{L}(r)\phi_{S}^{(L)}(t|r)\phi_{L,q}(t|r)\phi_{N,q}(t|r)f_{r,q}(r)dr \bigg] t^{-1}dt \nonumber \\ 
&- \dfrac{1}{\pi} \int_{0}^{\infty} Im\bigg[e^{-jt\theta_e}\phi_{D,q}(t) \int_{0}^{R}p_{N}(r)\phi_{S}^{(N)}(t|r)\phi_{L,q}(t|r)\phi_{N,q}(t|r)f_{r,q}(r)dr \bigg] t^{-1}dt \label{eqExpCDF} 
\end{align}
\setcounter{equation}{22}
\begin{align}
\mu_{C,q} = & \frac{B}{\ln(2)}\Bigg[\int_0^{\infty} g(t) \phi_{D,q} (jt)\int_{0}^{R}p_L(r)\Big(1-\phi_{S}^{(L)} \big(jt|r\big)\Big)\phi_{L,q} \big(jt|r\big)\phi_{N,q} \big(jt|r\big)f_{r,q}(r) dr dt \nonumber \\ 
&+ \int_0^{\infty} g(t) \phi_{D,q} (jt)\int_{0}^{R}p_N(r)\Big(1-\phi_{S}^{(N)} \big(jt|r\big)\Big)\phi_{L,q} \big(jt|r\big)\phi_{N,q} \big(jt|r\big)f_{r,q}(r) dr dt\Bigg]
    \label{eqCapAvg}
\end{align}

\hrulefill

\end{figure*}

\section{Analytical Results}
\label{anres}

In this section, the index $q=\{1,2\}$ is used to denote expressions related to a street UE and to a crossroad UE respectively.

\subsection{Preliminaries: characteristic functions}

A few characteristic functions (CF) are first introduced since they will be necessary to compute the performance metrics. All these CFs are conditioned on $r=d_{i^* j^*}$, the distance between the typical UE and its serving BS.

\setcounter{equation}{13}

\subsubsection{Fading models}
\begin{itemize}
\item[] All results of this section are developed to be applicable for any fading distribution. For this reason, they are systematically expressed as functions of the general CF of the fading model, denoted by $\phi_F(\cdot)$. The distributions that are selected as particular cases to generate the numerical results of section IV are provided below:
\begin{itemize}
    \item Rice fading is chosen for BSs in the user street(s). The CF of this fading model is given by Lemma 4.6 in \cite{francoisthesis}: 
    \begin{equation}
        \phi_F\big(t;K\big) = \dfrac{K+1}{K+1-jt} \exp\bigg[\dfrac{Kjt}{K+1-jt} \bigg]
    \end{equation}
    where $K$ is the K-factor of the distribution.
    \item Rayleigh fading is considered for BSs interfering via diffraction. The associated CF can be expressed as $\phi_F(t;K=0)$.
\end{itemize}

\end{itemize}

\subsubsection{CF of the useful received power}
\begin{itemize}
\item[] The CF of the useful received power is given by 
    \begin{equation}
        \phi_{S}^{(p)}(t|r) = \phi_F\Big(P_B\kappa_p^{-1}\sqrt{r^2 +\Delta H^2}^{-\alpha_p}t;K\Big) 
        \label{eqcharfunS}
    \end{equation}
where $p= \{L,N\}$ depending on whether the serving BS is in LOS or NLOS conditions.
 \begin{proof}
    the result comes from the definition of the useful power, given by \eqref{PMC11} or \eqref{PMC12}.
\end{proof}
\end{itemize}
\subsubsection{CFs of the interference affecting street users}

\begin{itemize}
\item[-]The CF of the LOS interference coming from the typical street is given by equation \eqref{eqcharfunL} next page. 
 \begin{proof}
    The proof is available in Appendix \ref{eqcharfunLProof}.
\end{proof}

\item[-] The CF of the NLOS interference coming from the typical street is given by equation \eqref{eqcharfunN}.
 \begin{proof}
    The proof follows the same approach as the proof of equation \eqref{eqcharfunL}.
\end{proof}

\item[-] The CF of the interference associated to first order diffraction is given by equation \eqref{eqcharfunD}.
 \begin{proof}
    The proof is available in Appendix \ref{eqcharfunDProof}.
\end{proof}
\end{itemize}
\subsubsection{CFs of the interference affecting crossroad users}
\begin{itemize}
\item[]The expressions derived for the street UE can be easily generalized to the crossroad UE. Taking into account the double presence of access points, the following expressions are obtained:

\setcounter{equation}{18}

\begin{minipage}[c]{0.45\linewidth}
 \begin{eqnarray}
  \phi_{L,2}(t|r) = \phi_{L,1}^2(t|r) \nonumber \\
  \phi_{N,2}(t|r) = \phi_{N,1}^2(t|r) \nonumber
 \end{eqnarray}
\end{minipage} \hfill
\begin{minipage}[c]{0.45\linewidth}
 \begin{eqnarray}
    \phi_{D,2}(t) =  \phi_{D,1}^2(t|r).
 \end{eqnarray}
\end{minipage}
\end{itemize}    
\subsection{Coverage probability}
The coverage probability is given by equation \eqref{eqCovprob} where $q=1$ for a street UE and $q=2$ for a crossroad UE.
\begin{proof}
    The proof is available in Appendix \ref{eqCovprobProof}.
\end{proof}
\subsection{Exposure distribution}
The CDF of the exposure can be expressed as equation \eqref{eqExpCDF}. 
\begin{proof}
    The developments follow the same reasoning as Appendix \ref{eqCovprobProof}.
\end{proof}

\subsection{Joint SIR-exposure distribution}
A lower bound (LB) for the joint SIR-exposure distribution can be expressed as:
\setcounter{equation}{21}
\begin{equation}
    \bar{F}_q(\theta_c,\theta_e) = \max\Big(0,P_{c,q}(\theta_c)+P_{e,q}(\theta_e)-1 \Big).
\label{eqJointMetric}
\end{equation}
 \begin{proof}
    This result comes from the application of Fréchet inequalities \cite{frechet1935generalisation}.
\end{proof}

\subsection{Average capacity}
The average user capacity is given by equation \eqref{eqCapAvg} where $g(t) = t^{-1} e^{-tW}$.
 \begin{proof}
    The proof consists in applying the lemma proposed in equation (2) of \cite{hamdi2010useful}. The other steps follows the same reasoning as Appendix \ref{eqCovprobProof}.
\end{proof}

\subsection{Average exposure}
The average exposure received at a street UE can be decomposed as 
\setcounter{equation}{23}
\begin{equation}
    \mu_{E,1} = \mu_{L,1} + \mu_{N,1} + \mu_{D,1}
    \label{eqMeanExp}
\end{equation}
 where the average exposures associated to the three categories of powers are given by
 \begin{align}
     \mu_{L,1} &= \int_{0}^{R} P_B \kappa_L^{-1} 2\lambda_B p_L(r) \sqrt{r^2+\Delta H^2}^{-\alpha_L} dr \label{eqMeanExpLOSStreet} \\
     \mu_{N,1} &= \int_{0}^{R} P_B \kappa_N^{-1} 2\lambda_B p_N(r) \sqrt{r^2+\Delta H^2}^{-\alpha_N} dr \label{eqMeanExpLOSStreet}\\
     \mu_{D,1} &= \int_0^{R} \int_{r_s}^{R} P_B \kappa_D^{-1} \big[ r+r'+qrr'\big]^{-\alpha_D} \; 4 \lambda_S \lambda_B drdr' 
     \label{eqMeanExpDiff}.
 \end{align}
 \begin{proof}
    The proof follows from the application of Campbell theorem \cite{baccelibook}.
\end{proof}

 The average exposure of a crossroad UE is then easily deduced: $\mu_{E,2} = 2\mu_{E,1}$.
 
\subsection{Generalization to an arbitrary UE}
All the above metrics can be generalized to the case of an arbitrary UE (possibly located in a crossroad or a single street with probabilities $\eta$ and $1-\eta$). Using the law of total probability, one has for each metric:
\begin{equation}
    m = (1- \eta) m_1 + \eta m_2
\end{equation}
where $m$ is the considered metric, $m_1$ is its expression for the street UE and $m_2$ its expression for the crossroad UE. The general variables $P_c (\theta_c)$, $P_e (\theta_e)$, $\bar{F}(\theta_c,\theta_e)$, $\mu_{E}$ and $\mu_C$ can hence be obtained in this manner.


\section{Numerical results obtained via stochastic geometry}
In order to generate the results of this section, a network of infinite size is considered: the network size $R$ is hence set to $+ \infty$ in the analytical expressions, and to $128km$ in the MC simulations (value sufficiently large to reproduce an infinite environment from the point of view of the centric user).
\label{numres}
\subsection{Impact of the propagation mechanisms}
\begin{figure}[]
    \centering
    \begin{subfigure}{0.4\textwidth}
    \includegraphics[width = \textwidth]{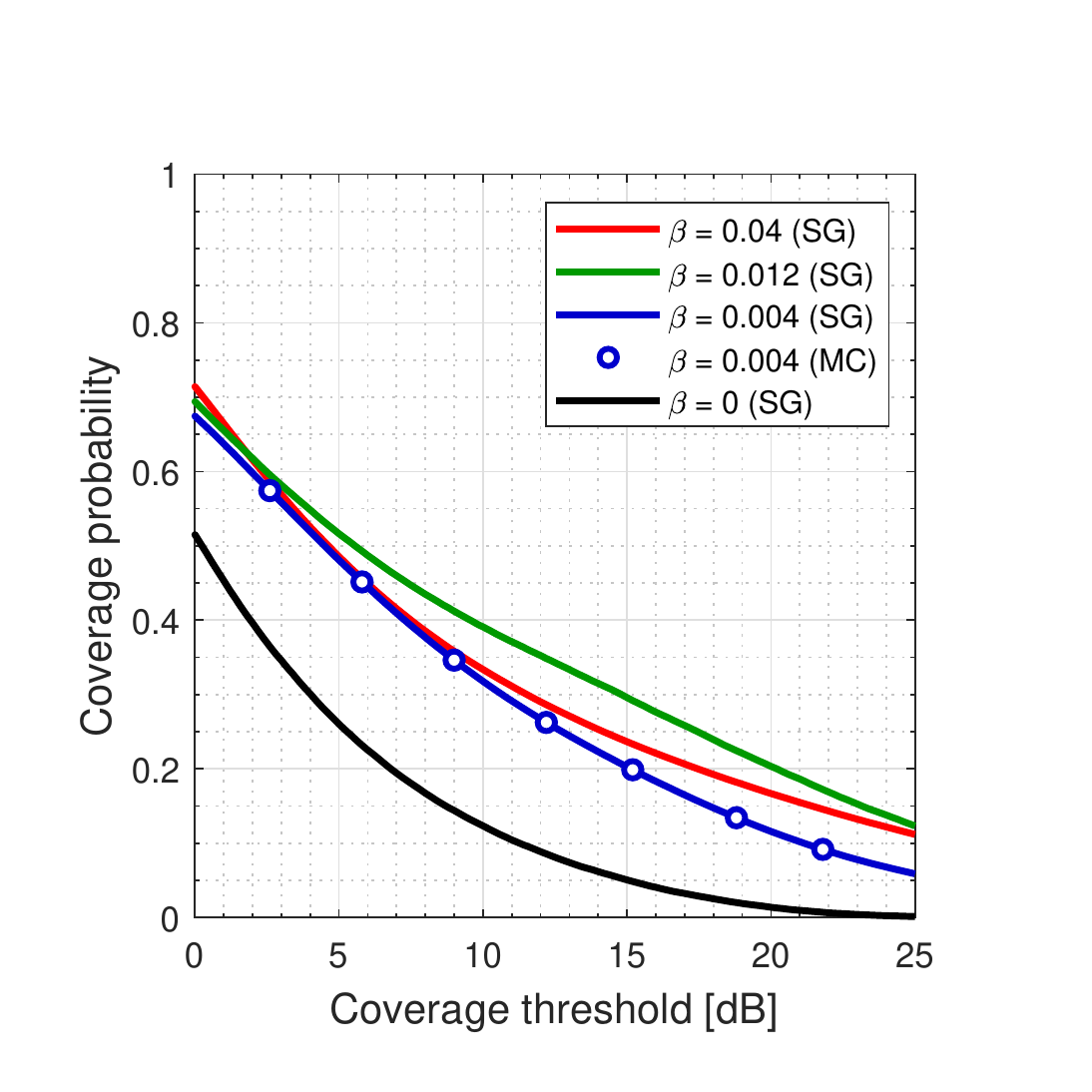}
    \caption{Coverage probability $P_c \left(\theta_c\right)$ for an arbitrary UE.} 
    \label{varbetaCov:a}
   \end{subfigure}
    \begin{subfigure}{0.4\textwidth}
    \includegraphics[width = \textwidth]{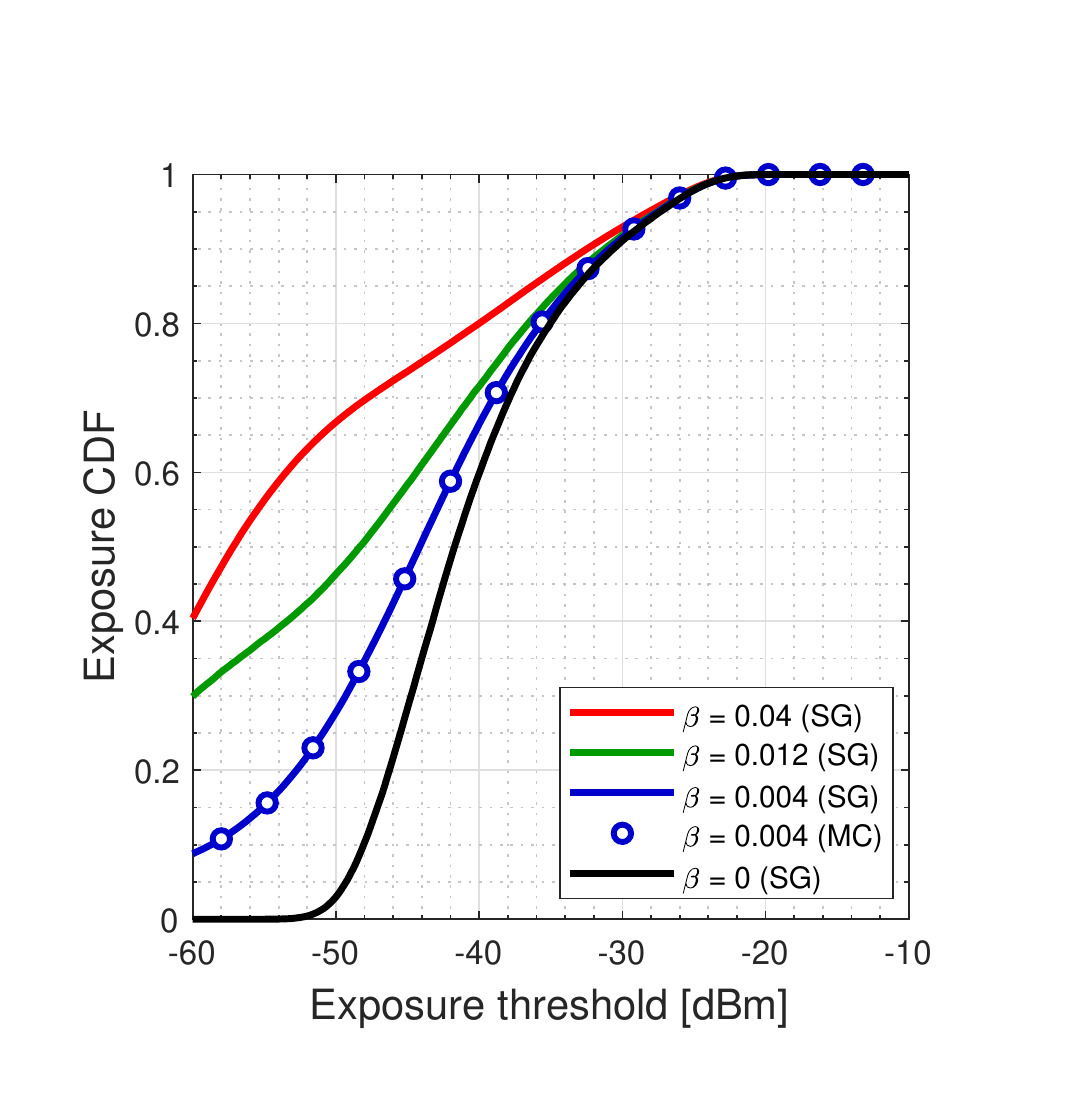}
    \caption{Exposure CDF $P_e \left(\theta_e\right)$ for an arbitrary UE.} \label{varbetaCov:b}
    \end{subfigure}
    \caption{Analysis of the impact of $\beta$. The parameters selected are $h_U=1.5m$, $h_B=6m$, $\lambda_S = 5/km$, $\lambda_B = 5/km$, $P_B = 1W$, $f=3.6GHz$, $\alpha_L = 1.7$, $\alpha_N = 2.5$, $\alpha_D = 3.5$, $\eta=0.1$, $\gamma = 1$ and $K=6$.}
    \label{varbetaCov}
\end{figure}

In this section, the roles of blockages and of diffraction are successively investigated. In order to study the impact of blockages, Figure \ref{varbetaCov} displays the coverage and the exposure evaluated for different LOS probabilities. For the studied cases, different values of the parameter $\beta$ (defining the LOS probability) are considered while the parameter $\gamma$ is fixed. The analytical results are validated using MC simulations and compared to the case without blockages ($\beta =0$). The evolution of the exposure (Figure \ref{varbetaCov:b}) is trivially explained by the presence of NLOS links, which increases with $\beta$ and reduces the received power.  Regarding the coverage (Figure \ref{varbetaCov:a}), two opposite scenarios can arise when introducing obstacles:
\begin{itemize}
    \item Scenario 1: the serving BS is in LOS and some of the interferers are in NLOS; this enhances the coverage probability when comparing to a case with no obstacle in the network.
    \item Scenario 2: the serving BS is NLOS while some interferers are still in LOS; this deteriorates the coverage probability.
\end{itemize}
These scenarios explain the impact of obstacles on the coverage. When progressively increasing the blockage probability up to $\beta = 0.012$, improvements in the coverage can be observed. These improvements therefore correspond to realizations associated to above scenario 1. By contrast, further increasing $\beta$ to 0.04 reduces the coverage. This decrease is in turn explained by a higher number of realizations corresponding to scenario 2 (due to the high value of $\beta$), which is detrimental to the coverage.

\begin{figure}
\centering
\begin{subfigure}{0.4\textwidth}
    \includegraphics[width=\textwidth]{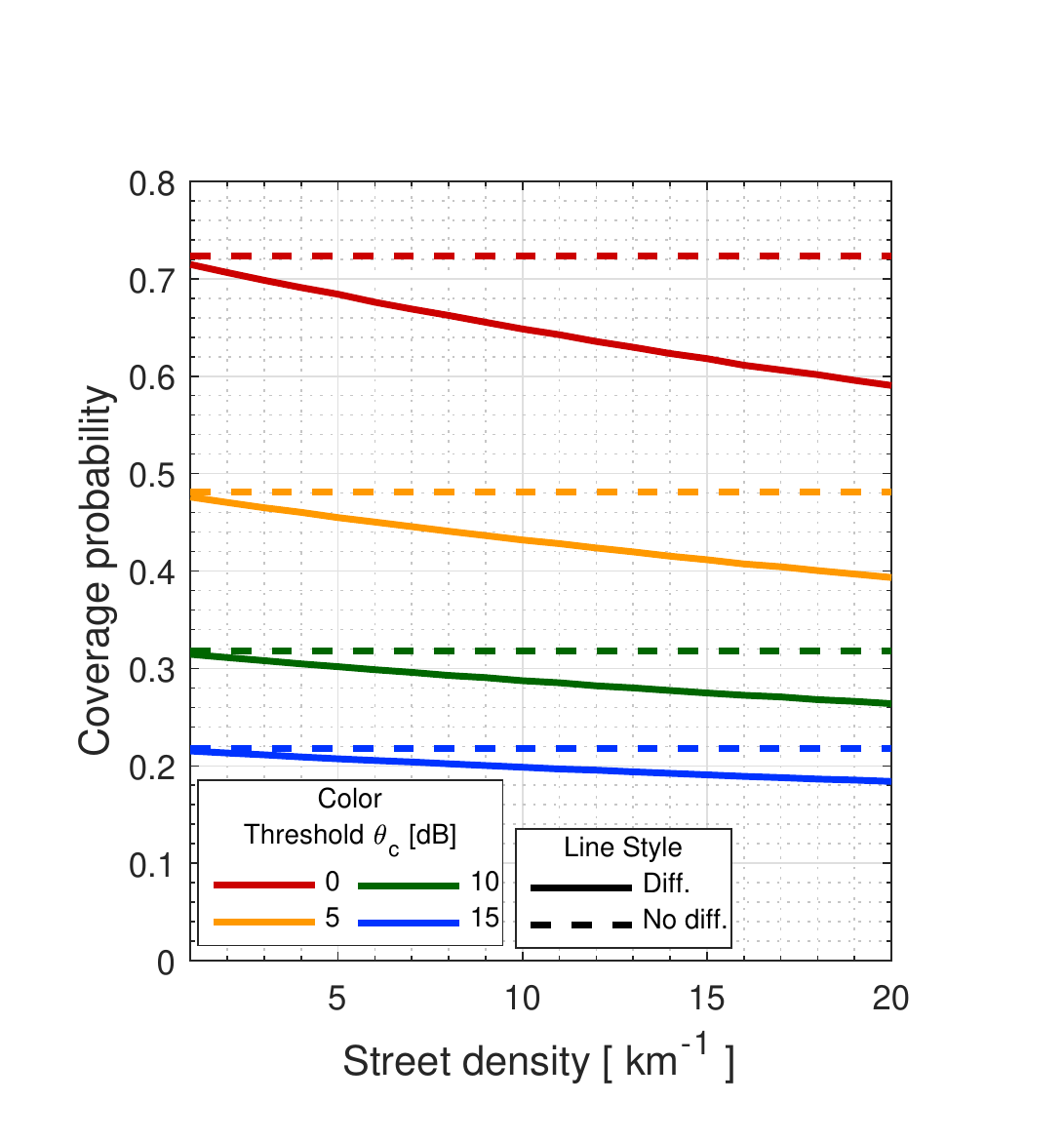}
    \caption{Impact of $\lambda_S$ on $P_{c,1}$.} \label{diff:a}
\end{subfigure}
\begin{subfigure}{0.4\textwidth}
    \includegraphics[width=\textwidth]{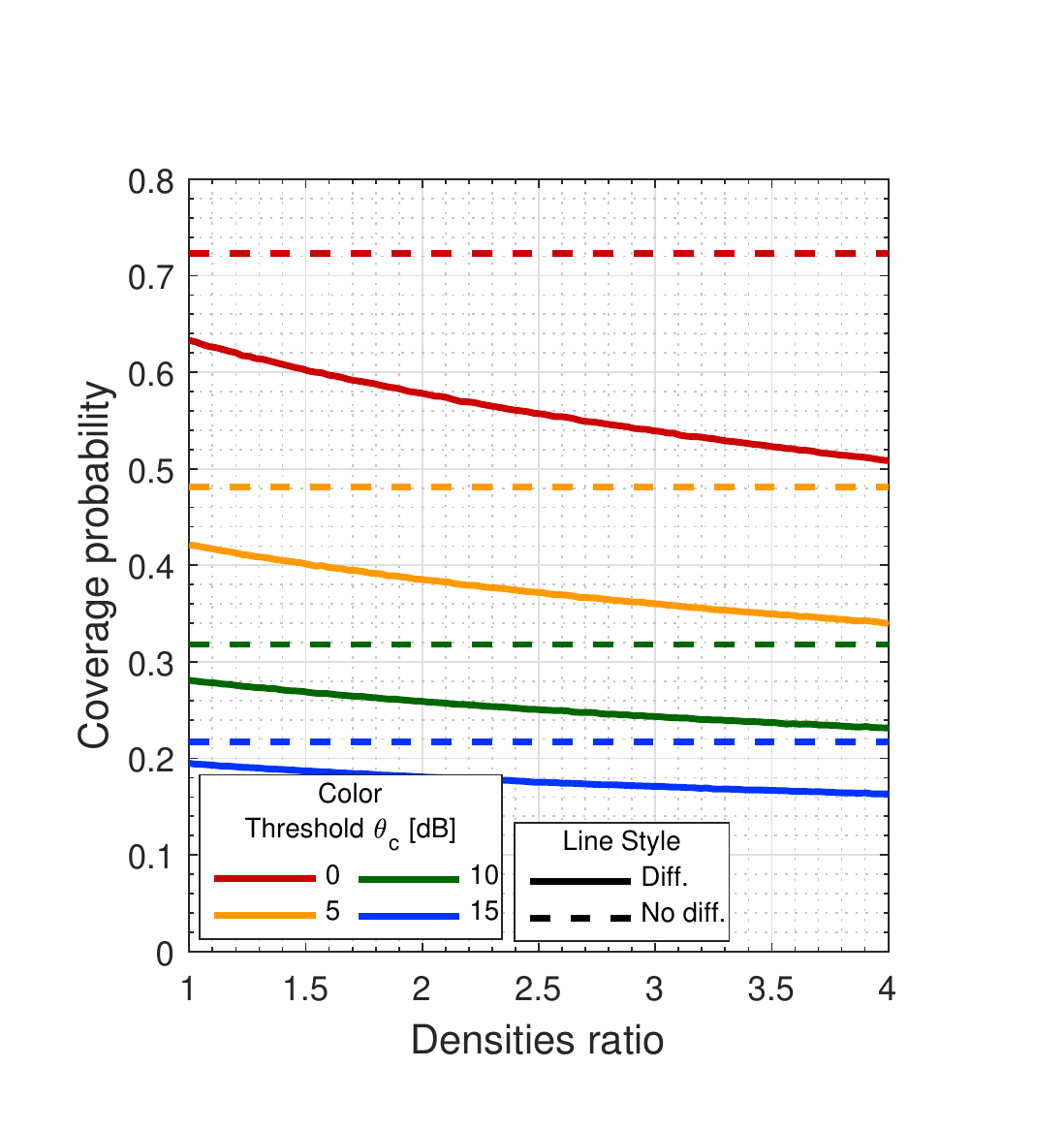}
    \caption{Impact of $\lambda_B^d/\lambda_B^t$ on $P_{c,1}$.} \label{diff:c}
\end{subfigure}
\caption{Impact of the BS and street densities. Unless stated otherwise, the selected parameters are $h_U=1.5m$, $h_B=6m$, $\lambda_S = 12.5km^{-1}$, $\lambda_B = 3km^{-1}$, $P_B = 1W$, $f=2GHz$, $\alpha_L = 1.7$, $\alpha_N = 2.5$, $\alpha_D = 2.5$, $\kappa_L = \kappa_N = \kappa_D = (4\pi f/c)^2$, $\beta = 0.04$, $\gamma = 1$ and $K=6$.} 
\label{diff}
\end{figure}

The graphs of Figure \ref{diff} quantify the influence of diffraction on the performance. For each of these graphs, coverage probabilities are shown for coverage thresholds of $0$, $5$, $10$ and $15$ dB. The continuous and dashed lines respectively represent the values obtained with and without diffracted signals taken into account. Figure \ref{diff:a} illustrates the evolution of these probabilities as function of the street density. Furthermore, it is also possible to show that the relative impact of diffraction can be more significant in street realizations characterized by a locally low BS density. For such cases, incident powers coming from BSs located outside the user street itself can proportionally have larger impacts on the coverage. Figure $\ref{diff:c}$ illustrates this statement by fixing the BS density $\lambda_B^t$ in the user street and by progressively increasing the BS density in the other streets, denoted by $\lambda_B^d$ in the legend. For a $5 dB$ threshold, the inclusion of diffracted signals results in a decrease in coverage of around 0.06 if the BS density is identical in all streets ($\lambda_B^d/\lambda_B^t=1$). When increasing the BS density in adjacent streets, this gap increases to 1.4 for $\lambda_B^d/\lambda_B^t=4$.

\subsection{Joint coverage and exposure distribution}
\begin{figure}
    \centering
\begin{subfigure}{0.4\textwidth}
    \includegraphics[trim=0.3cm 0 2cm 0,scale=0.65]{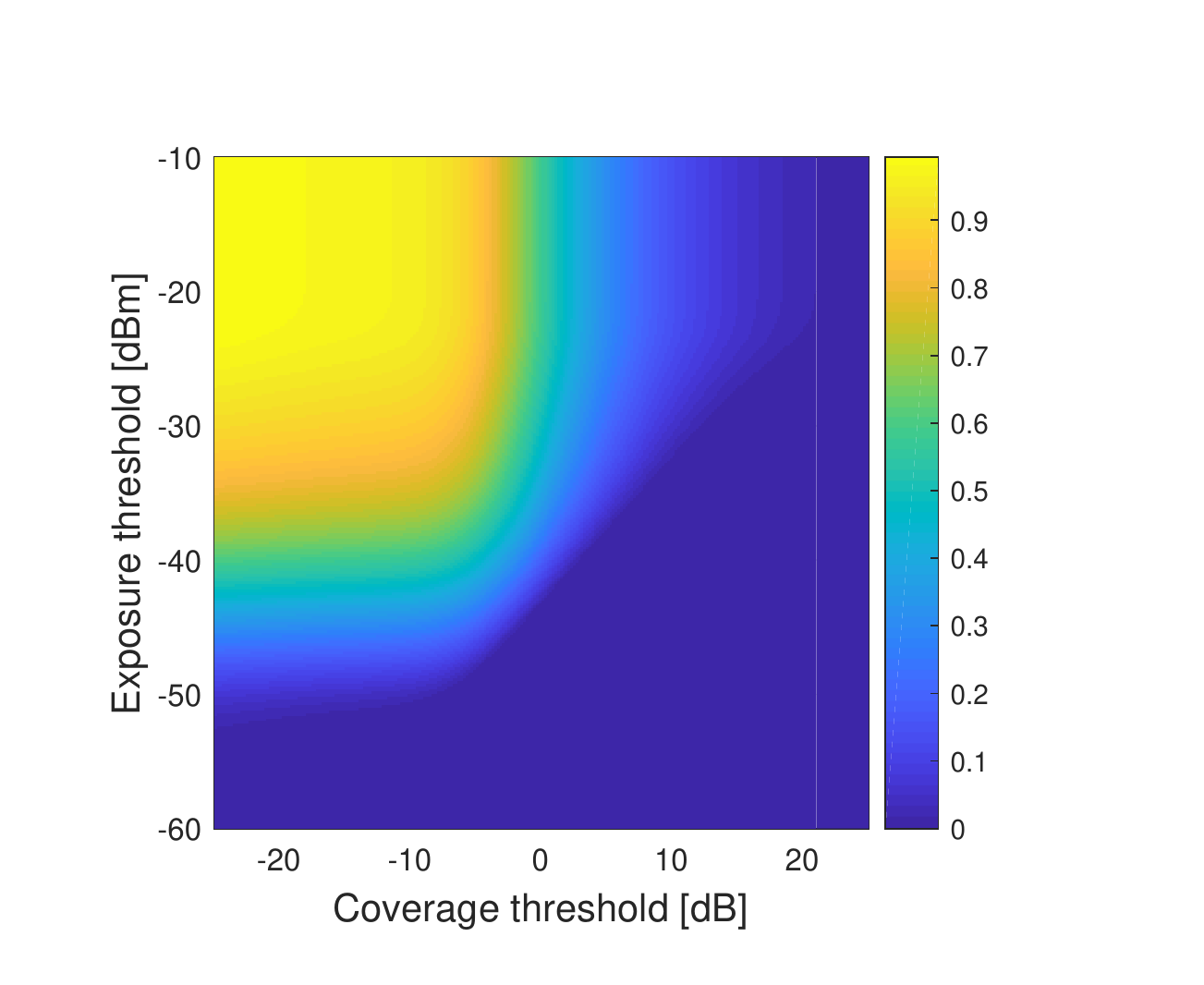}
    \caption{Lower bound on joint metric $F(\theta_c,\theta_e)$ for an arbitrary UE.} \label{jointmetricfig:a}
\end{subfigure}
\begin{subfigure}{0.4\textwidth}
    \includegraphics[width=\textwidth]{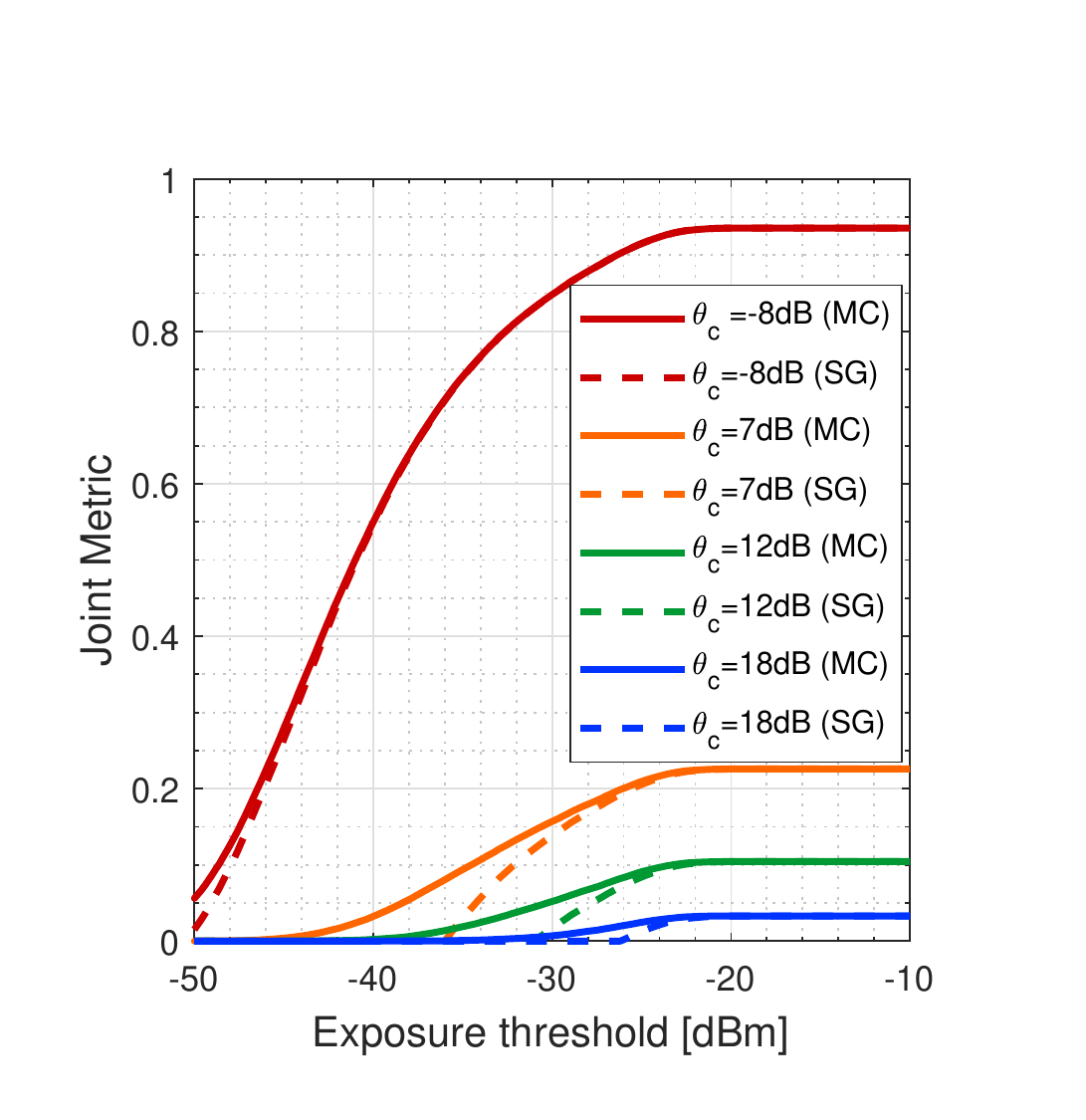}
    \caption{Assesment of the SG lower bound for fixed values of $\theta_c$.} \label{jointmetricfig:b}
\end{subfigure}
\caption{Analysis of the joint metric. The parameters selected are $h_U=1.5m$, $h_B=6m$, $\lambda_S = 5km^{-1}$, $\lambda_B = 5km^{-1}$, $P_B = 1W$, $f=3.6GHz$, $\alpha_L = 1.7$, $\alpha_N = 2.5$, $\alpha_D = 3.5$, $\kappa_L = \kappa_N = \kappa_D = (4\pi f/c)^2$, $\eta=0.1$, $\beta = 0.0004$, $\gamma = 1$ and $K=6$.} 
\label{jointmetricfig}
\end{figure}

Figure \ref{jointmetricfig} illustrates the joint coverage and exposure distribution. The SG lower bound of \eqref{eqJointMetric} is represented as function of the two thresholds $\theta_c$ and $\theta_e$ in Figure  \ref{jointmetricfig}(a). This bound is then compared to the MC simulations in Figure \ref{jointmetricfig}(b) for fixed values of $\theta_c$. One can observe on this graph that the probability of obtaining a low exposure decreases as the coverage threshold increases. This observation is due to the useful received power $S$, present in both SINR and exposure definitions. This power will statistically be high for users in good coverage, which increases for these users the probability to experience a high exposure. Some trade-offs are therefore necessary in the coverage and exposure requirements.

\subsection{User type comparison}
\begin{figure}
\centering
\begin{subfigure}{0.4\textwidth}
    \includegraphics[width=\textwidth]{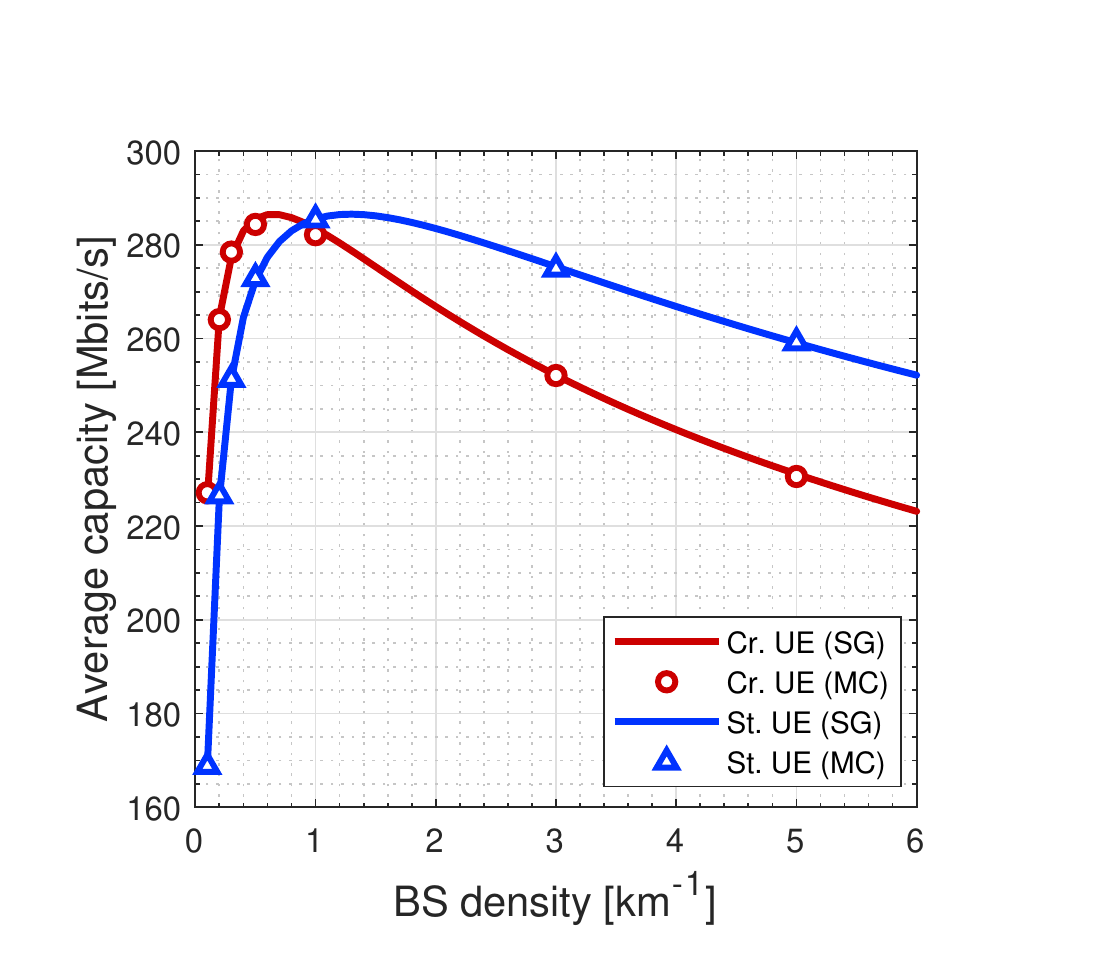}
    \caption{Average UE capacity for $\alpha_L = 2$, $\alpha_N = 2.25$ and $W=-93dBm$.} \label{avguecap:a}
\end{subfigure}
\begin{subfigure}{0.4\textwidth}
    \includegraphics[width=\textwidth]{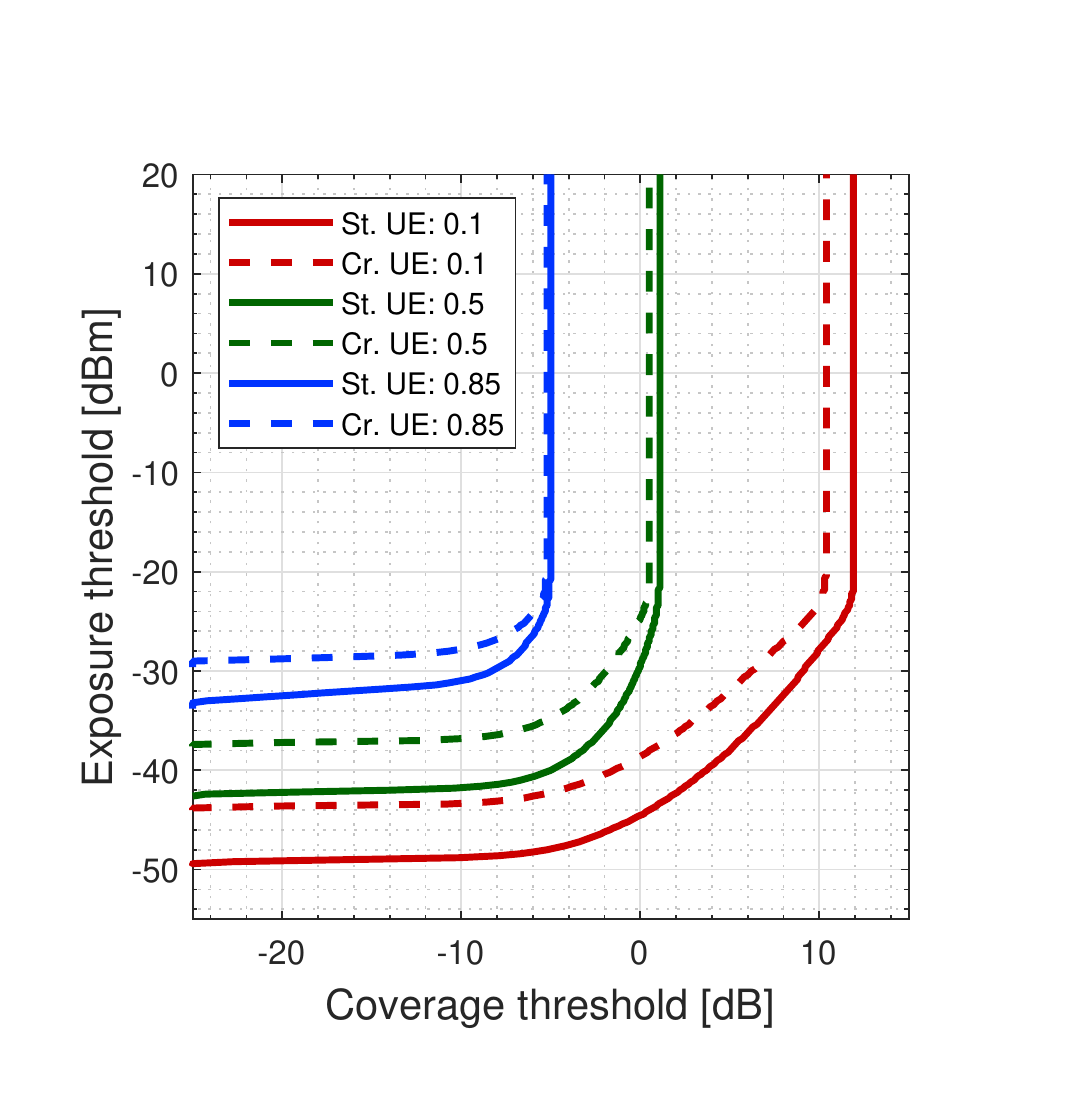}
    \caption{Isocurves of the joint distribution for $F(\theta_c,\theta_e) \in \{0.1,0.5,0.85\}$, $\alpha_L = 2.5$, $\alpha_N = 2.7$ and $W=0W$.} \label{avguecap:b}
\end{subfigure}
\caption{Analysis of the differences between street (st.) UE and crossroad (cr.) UE. The selected parameters are $h_U=1.5m$, $h_B=4.5m$, $\lambda_S = 5km^{-1}$, $\lambda_B = 5km^{-1}$ (for figure (b)), $f=3.6GHz$, $\alpha_D = 3.5$, $\kappa_L = \kappa_N = \kappa_D = (4\pi f/c)^2$, $\eta=0.1$, $\beta = 0.004$, $\gamma = 1$, $B=100MHz$ and $K=1$.} 
\label{avguecap}
\end{figure}

Figure \ref{avguecap} compares the performance achievable for a street (st.) UE and for a crossroad (cr.) UE. Figure \ref{avguecap:a} shows the average user capacity as a function of the BS density. Independently of the UE type, one can note the existence of an optimal value for this quantity, as already shown in previous works \cite{optimumrate}. For a low BS density, the signal-to-noise ratio takes low values, to the detriment of the capacity. As the BS density progressively increases, the serving BS statistically becomes closer to the UE, which increases the useful received power. The densification hence enables to increase the average capacity, up to some optimum $\lambda_B = \lambda_B^{opt}$. Beyond this maximum, the improvements in useful power no longer compensate the increase of interference, resulting in a decreasing capacity. One will note that the optimal capacity associated to street users is reached for a higher BS density compared to the optimal capacity of crossroad users. This difference is due to the presence of BSs which is in average doubled for crossroad users (due to the two streets forming the intersection). In Figure \ref{avguecap:b}, isocurves of the joint SIR-exposure metric are analysed. Each curve represents the set of thresholds $(\theta_c,\theta_e)$ satisfied with probability $p \in \{0.1,0.5,0.85\}$, for a street or a crossroad user. One can here observe that the achievable performance of the street user is higher than the crossroad user. One justification of this difference comes from the exposure of the crossroad user, which is in average doubled compared to the street user. At a given coverage requirement, the probability for the crossroad user to satisfy an exposure threshold is therefore reduced compared to the street user. 

\subsection{Optimal UE capacity}
\begin{figure}
\centering
\begin{subfigure}{0.4\textwidth}
    \includegraphics[width=\textwidth]{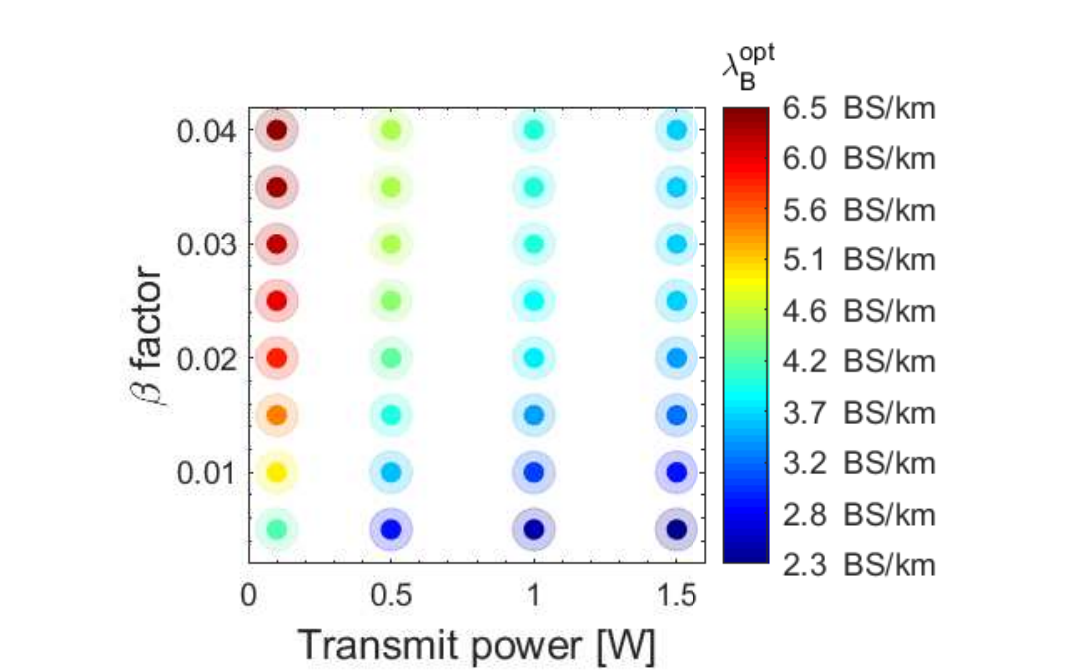}
    \caption{Optimal BS density $\lambda_B^{opt}$.} \label{optuecap:a}
\end{subfigure}
\begin{subfigure}{0.4\textwidth}
    \includegraphics[width=\textwidth]{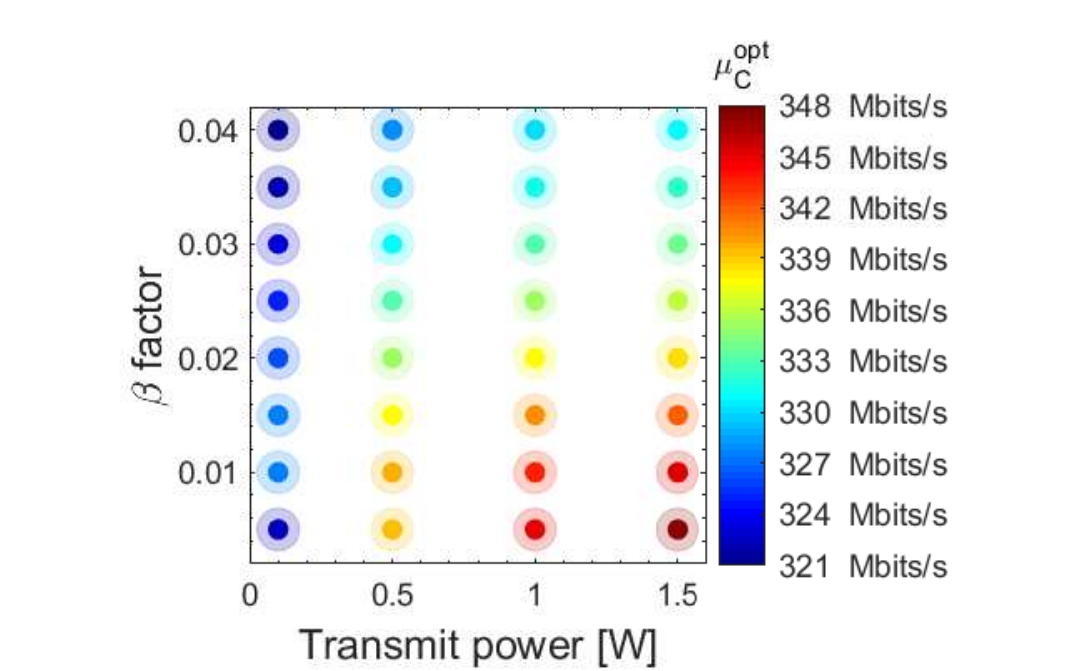}
    \caption{Associated maximal average capacity $\mu_C^{opt}$.} \label{optuecap:b}
\end{subfigure}
\caption{Analysis of the UE capacity as function of $\beta$ and $P_B$. The parameters selected are $h_U=1.5m$, $h_B=4.5m$, $\lambda_S = 5km^{-1}$, $\lambda_B = 5km^{-1}$, $f=3.6GHz$, $\alpha_L = 2.5$, $\alpha_N = 2.75$, $\alpha_D = 3.5$, $\eta=0.1$, $\beta = 0.004$, $\kappa_L = \kappa_N = \kappa_D = (4\pi f/c)^2$, $\gamma = 1$, $B=100MHz$, $W=-93dBm$ and $K=1$.} 
\label{optuecap}
\end{figure}

Figure \ref{optuecap:a} quantifies the variation of the BS density maximizing the mean capacity as a function of variables $\beta$ and $P_B$. This optimal density increases with the blockage probability factor $\beta$. Such an increase enables to compensate for the impact of $\beta$ on the probability of obtaining a NLOS serving BS. Regarding the transmit power, the SNR values are by definition proportional to $P_B$ and therefore decrease when its value is reduced. To compensate for this effect, the optimal BS density therefore increases to lower the path affecting the serving BS. The corresponding maximal capacities are shown in Figure \ref{optuecap:b}. One can note that even if the BS density is optimized for each pair $(P_B,\beta)$, the cases of low blockage probability and high transmit power lead to the highest optimal capacity values.  

\section{Ray tracing-based validation}
\label{compRT}
\subsection{RT network topology}
In the framework of the RT simulations, the studied environments are identical to the SG model, up to the following modifications:
\begin{itemize}
    \item Streets are given a non-zero width $w_S$.
    \item Users and BSs and are located on sidewalks. In a given street, the sidewalk on which each BS is located is randomly chosen. For the sake of practical RT implementation, those are placed at a slight distance $d_{B,BU}$ from the buildings walls.
    \item In order to model obstacles, parallelepipeds (representing cars) are generated with density $\lambda_O$ in the typical street(s). While the blockage model of the SG framework considers independent blockage probabilities, these obstacles here induce spatial correlations in the blockages experienced by adjacent BSs.
\end{itemize}

\subsection{RT power computation}
The UCLouvain RT software was used in the framework of this study \cite{RTTool}. Two propagation mechanisms are considered by this tool: reflection and diffraction. Also, only outdoor to outdoor scenarios are considered: the potential propagation of waves via building penetration is not modelled. To keep the computational time in a reasonable range, at most two interactions are considered between a BS and a UE. In addition, diffraction is only taken into account for the last interaction. These hypotheses seem reasonable as an electromagnetic wave which undergoes more than two interactions or more than one diffraction is strongly attenuated.
The computation of  the electrical fields for each propagation mechanism is detailed below. One will denote the distance between positions X and Y by $d_{XY}$.

\subsubsection{Line of sight}
Let $\overline{e}_0$ be the unit vector aligned with the transmitted electric field. The LOS field received from a BS located in $B$ at an observation point $P$ is given by
\begin{equation}
    \overline{e}_{LOS}\left(P\right) =  \overline{e}_{0} \frac{e^{-jkd_{BP}}}{d_{BP}},
    \label{elos}
\end{equation}

where $k$ is the wave number.

\subsubsection{Reflection}
Let $Q$ be a reflection point. Using Fresnel dyadic coefficients $\overline{\overline{R}}$ (computed as in \cite{oestges}), the reflected field $\overline{e}_{ref}\left(P\right)$ at observation point $P$ is given by
\begin{equation}
    \overline{e}_{ref}\left(P\right) = \overline{\overline{R}} \cdot \overline{e}_{LOS}\left(Q\right) \frac{d_{BQ}}{d_{BQ}+d_{QP}} e^{-jkd_{QP}}.
    \label{eref}
\end{equation}

\subsubsection{Diffraction}
Let $S$ be a diffraction point. Using uniform theory of diffraction (UTD), the reflected field $\overline{e}_{diff}\left(P\right)$ at observation point $P$ is given by 
\begin{equation}
    \overline{e}_{diff}\left(P\right) = \overline{\overline{D}} \cdot \overline{e}_{LOS}\left(S\right) \sqrt{\frac{d_{BS}}{d_{SP}\left(d_{BS}+d_{SP}\right)}} e^{-jkd_{SP}},
    \label{ediff}
\end{equation}
where $\overline{\overline{D}}$ are the Guevara's coefficients preferred to the usual UTD dyadic coefficients for their reciprocity regarding the ray's incidence \cite{tami}. 

\subsubsection{Received power}
Finally, since $\overline{e}_{0}$ is normalized at the emission, the received power at a typical user location $i^*$ from a given BS $j$ is given by
\begin{align}
          &P_{i^*j}^{(RT)} =  P_B \Big(\dfrac{4\pi f}{c}\Big)^{-2} ||\overline{e}_{tot}\cdot\overline{e}_{tot}^H|| \\
          &\text{with} \quad  \overline{e}_{tot} = \sum_{n=1}^{N_{paths}} \overline{e}_{n,j} 
          \label{rxpow}
\end{align}
where 
\begin{itemize}
    \item $N_{paths}$ is the number of paths coming from this BS and arriving at the typical user under the considered hypotheses.
    \item $P_B$ is the transmit effective isotropic radiated power.
    \item $c$ is the speed of light.
    \item $\overline{e}_{n,j}$ are the electric fields at $i^*$ computed by means of equations \eqref{elos}, \eqref{eref} and \eqref{ediff}.
\end{itemize} 
\subsection{Methodology to fix the parameters}
\label{rtnettop}
In order to compare the SG and RT approaches, the parameters of the two frameworks should be coherently selected. Table \ref{table_param} summarizes these macroparameters.
\begin{table}[H]
\renewcommand{\arraystretch}{1.2}
\centering
\small
\begin{tabular}{ |l| } 
  \hline
  \textbf{Parameters common to the RT and SG frameworks} \\
  \hline
  $R$, $h_U$, $h_B$, $r_s$, $\lambda_S$, $\lambda_B$, $P_B$, $B$ and $f$\\ 
  \hline
  \textbf{Parameters specific to the RT framework} \\
  \hline
  $w_S$, $d_{B,BU}$ and $\lambda_O$.   \\ 
  Ground permittivity: $15-1.50j$ (cfr. p. 66-67 of \cite{oestges})\\ 
  Building permittivity: $5.3-0.42j$ (cfr. p. 66-67 \cite{oestges}) \\ 
  \hline
  \textbf{Parameters specific to the SG framework} \\
  \hline
  $\beta$, $\gamma$, $\alpha_L$, $\alpha_N$, $\alpha_D$, $\kappa_L$, $\kappa_N$, $\kappa_D$, $\eta$, fading models  \\ 
  \hline
\end{tabular}
\caption{Parameters employed in the SG and RT frameworks.}
\label{table_param}
\end{table}

In the considered case study, the parameters of the first and second line of table \ref{table_param} were set to $R=2km$, $h_U=1.5m$, $h_B=6m$, $r_s = 1m$, $\lambda_S = 5km^{-1}$, $\lambda_B = 5km^{-1}$, $P_B = 1W$, $B=100MHz$, $f=3.6GHz$, $w_S=35m$, $d_{B,BU}=5m$ and $\lambda_O=20km^{-1}$. We also set $\kappa_L = \kappa_N = \kappa_D = (4\pi f/c)^2$. From a sufficient amount of RT simulations, we are then able to estimate the other SG parameters based on the statistics of the RT results. The employed methodology consists of the following steps: 
\begin{itemize}
    \item[-] RT simulations are performed for a large number of realizations of the MPLP. At each iteration, the received powers $P_{i^*j}^{(RT)}$ at the typical UE are computed using equation \eqref{rxpow}. The propagation condition (LOS or NLOS) of each link is also assessed.
    \item[-] Based on the statistics of these received powers, the parameters $\beta$, $\gamma$, $\alpha_L$ and $\alpha_N$ are computed using a mean square error method.
    \item[-] The corresponding fading distribution is then deduced by dividing the received powers by the path losses estimated at the previous step. An analytical fading model is then selected to approximate this empirical distribution. 
    \item[-] The parameter $\eta$ is independently determined by computing the percentage of users located at crossroads in the RT simulations.
\end{itemize}
The values obtained for the SG parameters were then $\beta = 0.004$, $\gamma = 0.85$, $\alpha_L = 1.66$, $\alpha_N = 1.93$ and $\eta=0.02$. The squared fading gains (i.e. fading powers) of the LOS and NLOS links were approximated with exponential distributions of rate parameters $1.66$ and $0.33$ respectively.


\subsection{Numerical results}
\begin{figure}
\centering
\begin{subfigure}{0.4\textwidth}
    \includegraphics[trim={0 0 0 1.5cm},width=\textwidth]{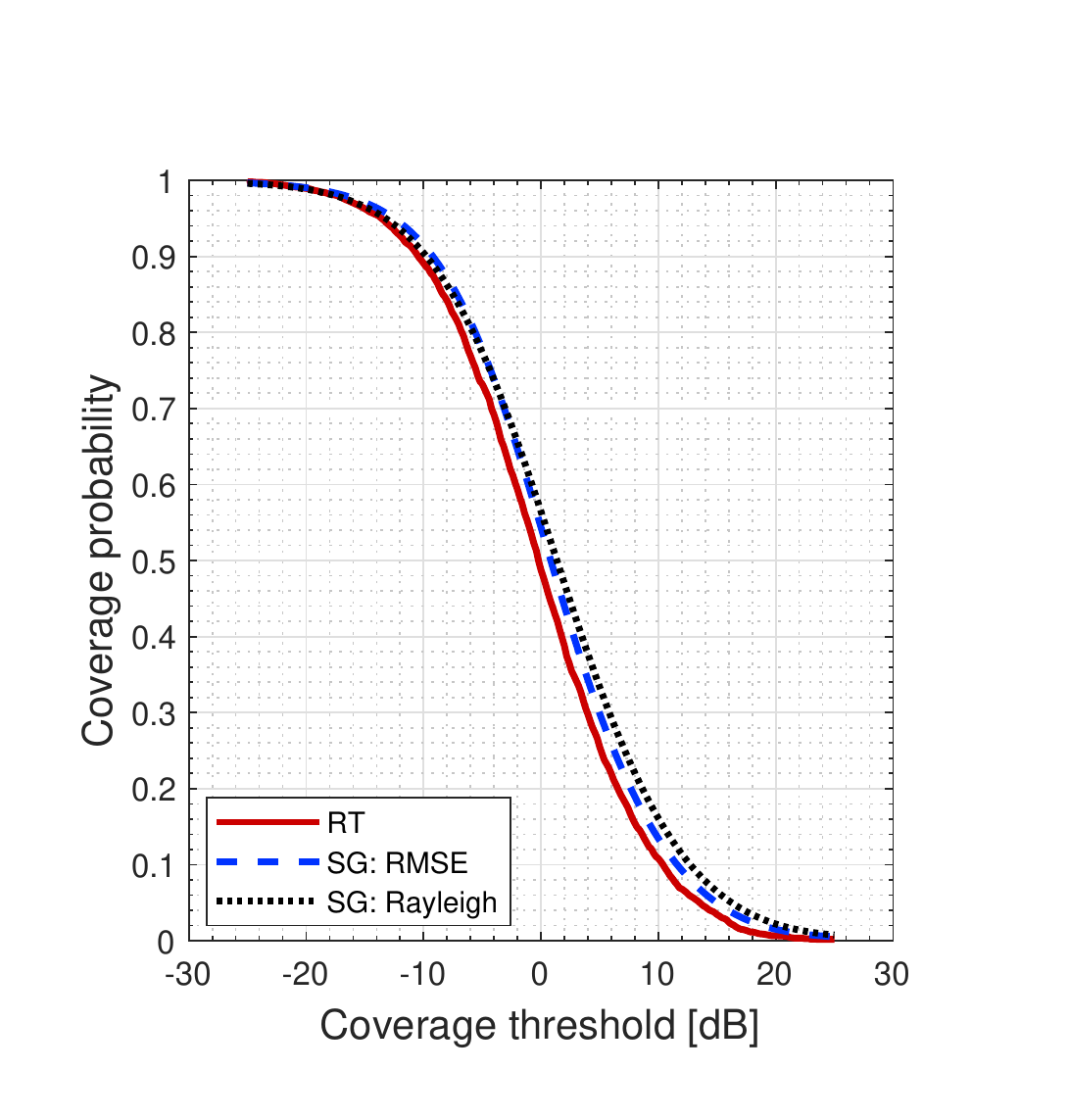}
    \caption{Coverage probability $P_c \left(\theta_c\right)$.} \label{comp:a}
\end{subfigure}
\par\vspace{0.19cm}
\begin{subfigure}{0.4\textwidth}
    \includegraphics[trim={0 0 0 0cm},width=\textwidth]{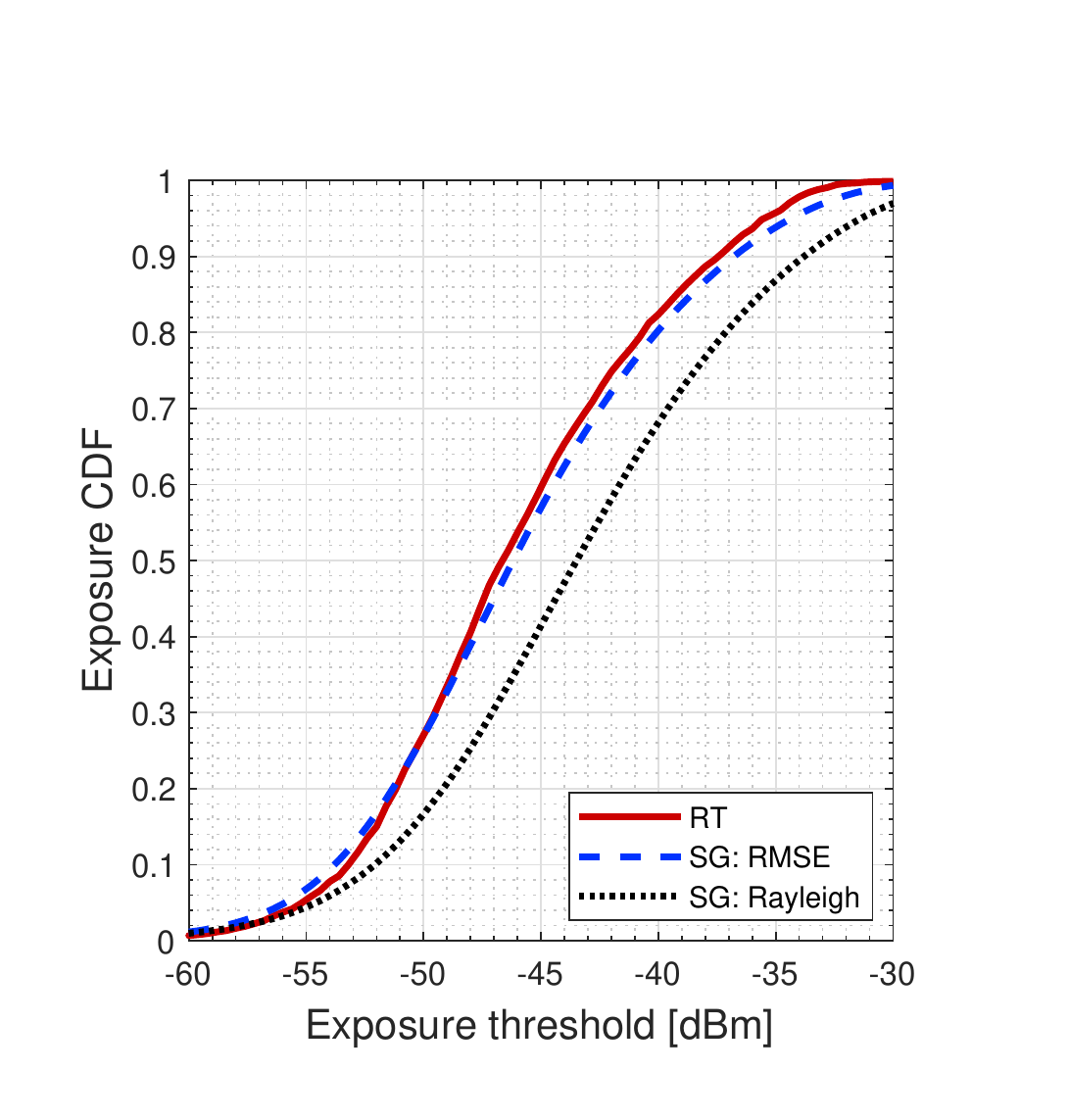}
    \caption{Exposure CDF $P_e \left(\theta_e\right)$.} \label{comp:b}
\end{subfigure}
\begin{subfigure}{0.4\textwidth}
    \includegraphics[width=\textwidth]{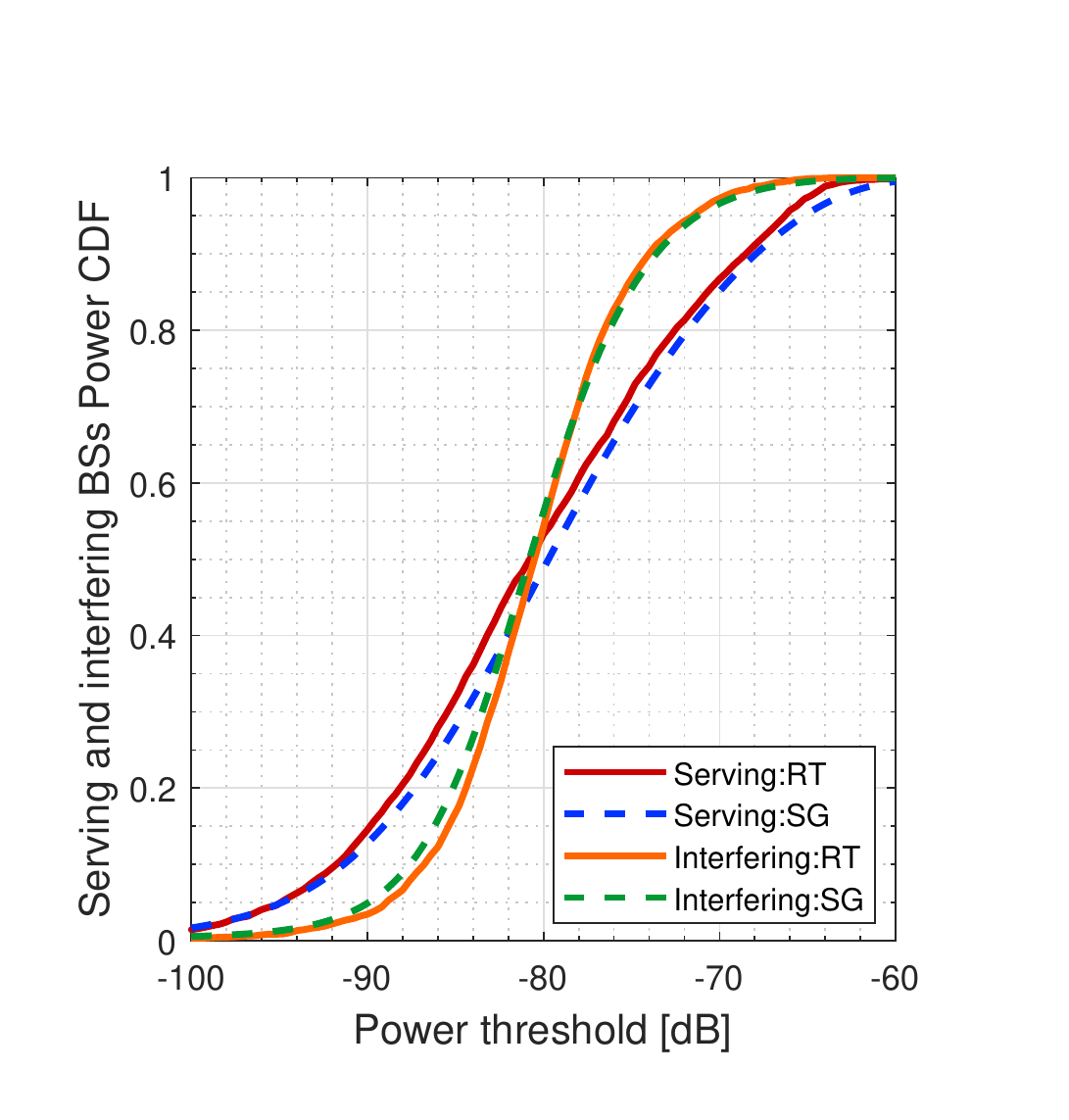}
    \caption{Received and interfering powers (RMSE only)} \label{comp:c}
\end{subfigure}
\caption{Comparison of SG and RT approaches.} 
\label{comp}
\end{figure}

Both frameworks were compared for the parameters obtained in the previous section. Figures \ref{comp:a} and \ref{comp:b} illustrate the coverage and exposure distributions. The slight differences between the RT and SG curves can be explained by means of Figure \ref{comp:c}. This figure shows the CDFs of the useful and interfering powers: $P\left(S>\theta_p\right)$ and $P\left(I_L +I_N +I_D >\theta_p\right)$. One can observe that the contribution of the serving BS in SG is slightly overestimated compared to the RT values. This results in coverage and exposure probabilities which are overestimated as well. Figures \ref{comp:a} and \ref{comp:b} also illustrate the results obtained when SG fading parameters are not optimized with respect to the RT tool. In that case, only the path loss parameters are optimized and the SG fading distribution is taken as normalized Rayleigh (i.e. with squared channel gains modelled via an exponential distribution of unit rate). One will note the loss of accuracy obtained when both path loss and fading aspects are not optimized.

\begin{figure}
    \centering
\begin{subfigure}{0.4\textwidth}
    \includegraphics[trim={0 0 0 1.2cm} ,width=\textwidth]{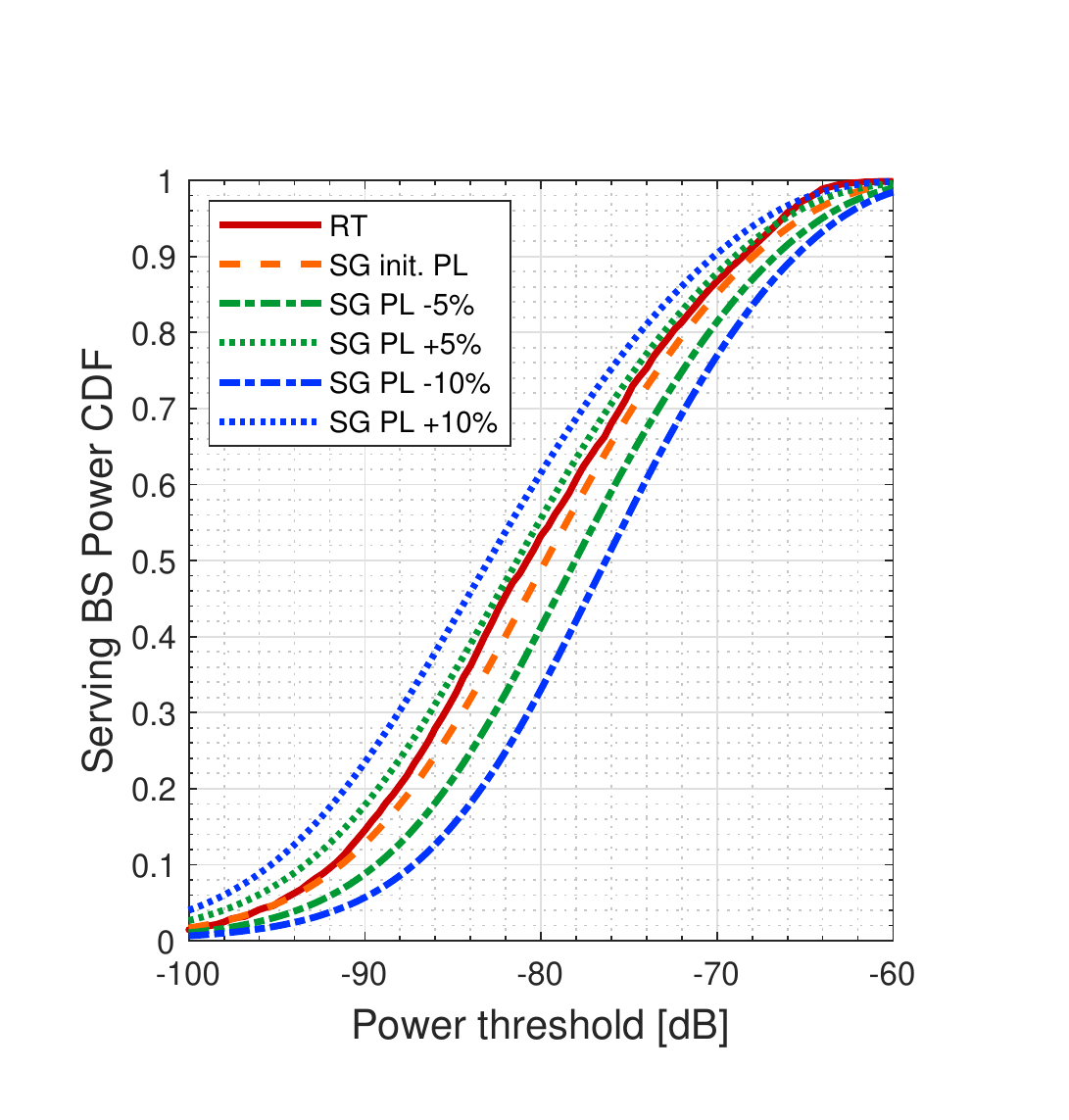}
            \caption[Network2]%
            {{\small Impact of PL exponents estimation errors on the CDF of the useful received power}}
            \label{sensitivity:a}
\end{subfigure}
\begin{subfigure}{0.4\textwidth}
    \includegraphics[trim={0 0 0 0.1cm} ,width=\textwidth]{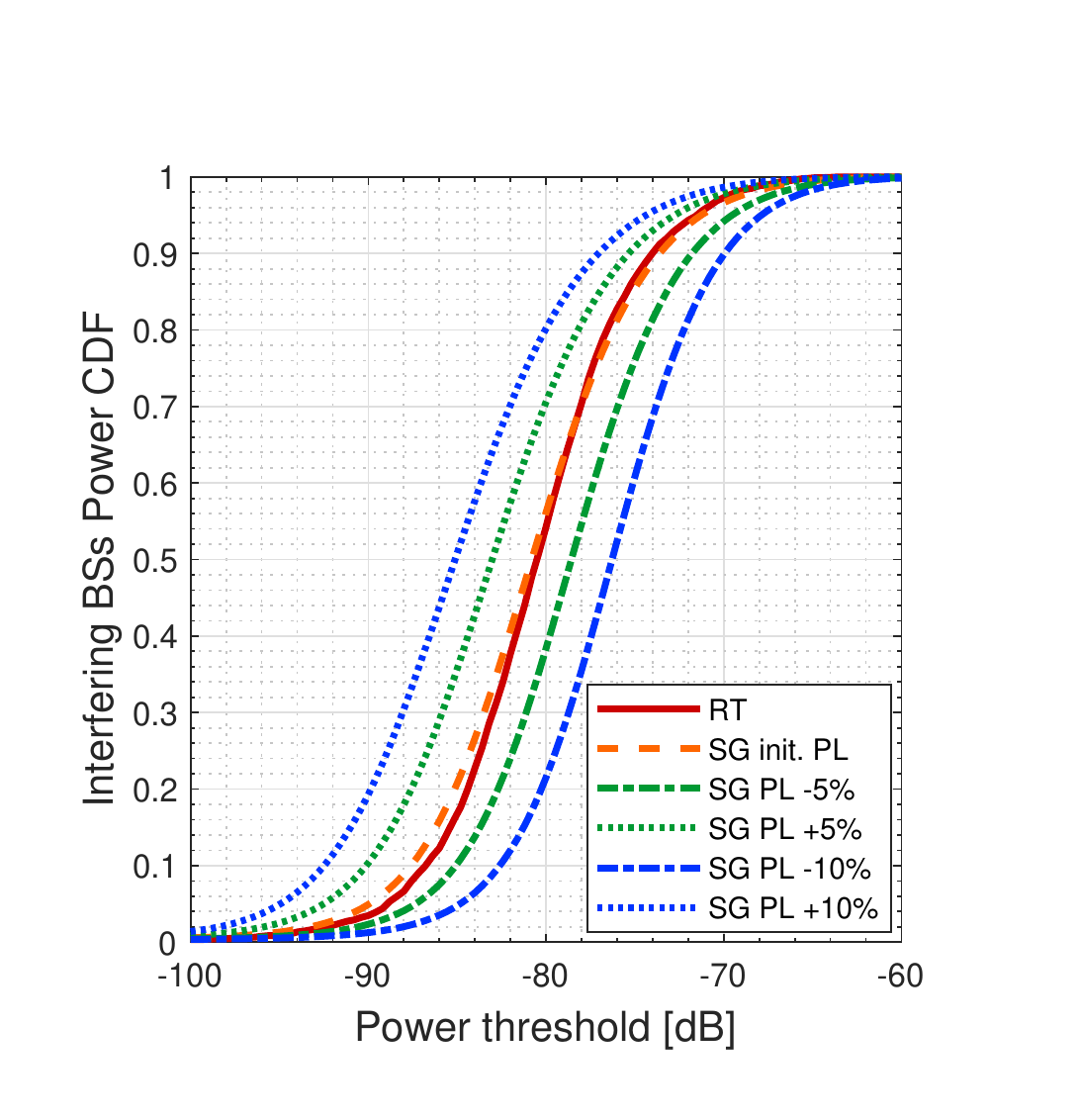}
            \caption[]%
            {{\small  Impact of PL estimation errors on the CDF of the aggregate interference}}    
            \label{sensitivity:b}
\end{subfigure}
\caption[ The average and standard deviation of critical parameters ]
        {\small Sensitivity analysis of the estimated PL exponents based on the CDFs of the useful received power $P\left(S>\theta_p\right)$ and of the aggregate interference $P\left(I_L +I_N +I_D >\theta_p\right)$.} 
\label{sensitivity}
\end{figure}

Figure \ref{sensitivity} illustrates the sensitivity of the SG model to potential errors in the estimated path loss exponents. Estimation errors of $5$ and $10$ percents are introduced. The impact on the distributions of the useful power and of the interference is shown in \ref{sensitivity:a} and \ref{sensitivity:b}. One can observe that these errors can lead to significant deviations, which highlights the importance of accurately extracting the parameters if SG is to be used. 

\begin{figure*}[t!]

\begin{align}
    \label{proof1}
    \phi_{L,1}(t|r)  &= \mathbb{E}_{\Phi_{L,r}}\left[\prod_{j \in \Phi_{L}(r) } \mathbb{E}_{\left|h_j\right|^2}\left[\exp{\left(  jtP_B\left|h_j\right|^2\kappa_L^{-1}\sqrt{r_j^2 + \Delta H^2}^{-\alpha_L} \right) }  \right]\right] \nonumber \\
    &\stackrel{(1)}{=} \exp \left(-2\lambda_{B}\int_{r}^R  p_L(r')\Bigg[1- \mathbb{E}_{\left|h\right|^2}\left[\exp{\left(  jtP_B\left|h\right|^2\kappa_L^{-1}\sqrt{r'^2 + \Delta H^2}^{-\alpha_L} \right) }  \right]\Bigg] dr' \right)   \nonumber\\
    &\stackrel{(2)}{=} \exp\Bigg[2\lambda_{B}\int_{r}^{R} p_{L}(r')\bigg[ \phi_F\Big(P_B\kappa_L^{-1}\sqrt{r'^2 +\Delta H^2}^{-\alpha_L}t;K\Big) - 1\bigg] dr' \Bigg]. \\
    \label{proof2}
    \phi_{D,1}(t|r)  &= \mathbb{E}_{\Psi_{VS}}\left[\mathbb{E}_{\Psi_{b,k}}\left[ \mathbb{E}_{\left|h_{jk}\right|^2}\left[ \exp{\left( \sum_{y_k \in \Phi_{VS}} \sum_{x_{jk} \in \Phi_{b,k}} jtP_B\kappa_D^{-1}\left|h_{jk}\right|^2\mathcal{D}_b(x_{jk},y_{k})^{-\alpha_D}\right) }  \right]  \right] \right] \nonumber\\
    &= \mathbb{E}_{\Phi_{VS}}\left[\prod_{y_k \in \Psi_{VS}} \mathbb{E}_{\Psi_{b,k}}\left[\prod_{x_{jk} \in \Psi_{b,k}} \mathbb{E}_{\left|h_{jk}\right|^2}\left[\exp{\left(  jtP_B\kappa_D^{-1}\left|h_{jk}\right|^2\mathcal{D}_b(x_{jk},y_{k})^{-\alpha_D} \right) }  \right]\right] \right] \nonumber\\
    &\stackrel{(1)}{=} \mathbb{E}_{\Phi_{VS}}\left[\prod_{y_k \in \Psi_{VS}} \mathbb{E}_{\Psi_{b,k}}\left[\prod_{x_{jk} \in \Psi_{b,k}} \Phi_F\Big(P_B\kappa_D^{-1}\mathcal{D}_b(x,y)^{-\alpha_D}t;0\Big) \right] \right] \nonumber\\
    &\stackrel{(2)}{=} \mathbb{E}_{\Phi_{VS}}\left[\prod_{y_k \in \Psi_{VS}}  \exp \left(-2\lambda_{B}\int_{r_s}^R  1- \Phi_F\Big(P_B\kappa_D^{-1}\mathcal{D}_b(x,y)^{-\alpha_D}t;0\Big) dx \right)\right] \nonumber \\ 
    &\stackrel{(3)}{=} \exp \Bigg[-2\lambda_{S}\int_{0}^R  1- \exp \Big[-2\lambda_{B}\int_{r_s}^R  1- \Phi_F\Big(P_B\kappa_D^{-1}\mathcal{D}_b(x,y)^{-\alpha_D}t;0\Big)  dx \Big]dy\Bigg] \\
    \label{proof3}
    P_{c,q}^{(p)}(\theta_c|r)  &\stackrel{(1)}{=} \mathbb{P}\left[\left.S-\theta_c\left(I_L +I_N +I_D\right)>0\right|r\right] \nonumber\\
    &\stackrel{(2)}{=} \dfrac{1}{2} + \dfrac{1}{\pi} \int_{0}^{\infty} Im\bigg[\phi_{S}^{(p)}(t|r)\phi_{L,q}(-\theta_c t|r) \phi_{N,q}(-\theta_c t|r)\phi_{D,q}(-\theta_c t) \bigg] t^{-1}dt.  \\
    \label{proof4}
    P_{c,q}(\theta_c)&\stackrel{(1)}{=} \int_{0}^{R}  \left(p_L(r)P_{c,q}^{(L)}(\theta_c|r) + p_N(r)P_{c,q}^{(N)}(\theta_c|r) \right)f_{r,q} (r)dt \nonumber\\  
    &\stackrel{(2)}{=} \dfrac{1}{2} + \dfrac{1}{\pi} \int_{0}^{\infty} Im\bigg[\phi_{D,q}(-\theta_c t) \int_{r_0}^{R}p_L(r)\phi_{S}^{(L)}(t|r)\phi_{L,q}(-\theta_c t|r)\phi_{N,q}(-\theta_c t|r)f_{r,q}(r)dr \bigg] t^{-1}dt \nonumber\\
    & ~\quad ~\quad+ \dfrac{1}{\pi} \int_{0}^{\infty} Im\bigg[\phi_{D,q}(-\theta_c t) \int_{r_0}^{R}p_N(r)\phi_{S}^{(N)}(t|r)\phi_{L,q}(-\theta_c t|r)\phi_{N,q}(-\theta_c t|r)f_{r,q}(r)dr \bigg] t^{-1}dt 
\end{align}

\hrulefill

\end{figure*}

\section{Conclusion}
\label{ccl}

In this work, a joint analysis of the data rate and exposure has been conducted for Manhattan environments. This analysis relied on several metrics, computed by means of SG and RT approaches. We have shown that both frameworks could yield similar performance for these metrics. An appropriate estimation of the propagation parameters is however a necessary condition to obtain such a correspondence. The RT validation therefore strengthens the relevance of using SG tools for Manhattan environments. In addition, it reinforces the idea of developing hybrid approaches: the RT tool has the advantage of relying on more physical backgrounds compared to the SG abstraction level. However, it is more time consuming than SG analyses, or than the equivalent MC simulations usually used to validate these SG approaches. When calculating the performance of users in large environments, one might consider combined techniques. On the one hand, power contributions from BSs close to the user could be computed via RT. On the other hand, power contributions from BSs located further away (with less impact) could be computed using these equivalent MC simulations. Such two-fold framework might provide interesting trade-offs in terms of accuracy and computational resources. Several commercial softwares have been built with this line of thought, but with more complex models (see for example \cite{winprop} combining COST and RT approaches). Our results might hence open the door to further simplifications.

Other work extensions could include the incorporation of beamforming schemes. Moreover, the RT tool could be improved by implementing additional propagation mechanisms (e.g. diffuse scattering). Finally, the analysis might be extended by proposing a comparison to a third framework (e.g. Winner type models).  

\appendices
  \section{Proof of Equation \eqref{eqcharfunL}: characteristic function for the LOS interfering links}
  \label{eqcharfunLProof}

  This proof is inspired by \cite{elsawy2016modeling}. Let $\Phi_{L,r}$ be the set of interfering BSs located in the typical street in LOS conditions located further than a distance $r$ from the centric user. We denote by $\left|h_j\right|^2$, the fading coefficients associated to a BS $j \in \Phi_{L,r}$ and $r_j$, its distance to the typical user. The characteristic function of the interference coming from $\Phi_{L,r}$ is given by \eqref{proof1} at the top of the page. In the developments of this equation, (1) comes is obtained by definition of the probability generating functional (PGFL) of $\Phi_{L,r}$; (2) comes from the definition of the characteristic function $\Phi_F(\cdot;\cdot)$.
  
  \section{Proof of Equation \eqref{eqcharfunD}: characteristic function of the diffracted links}
  \label{eqcharfunDProof}
  For this proof, the following notations are introduced:
  \begin{itemize}
      \item $\Psi_{VS}$ it the 1D PPP of the intersections of the vertical streets with the typical UE avenue.
      \item $y_k$ is the distance between the typical UE and the crossroad with vertical street $k\in \Psi_{VS}$. 
      \item $\Psi_{b,k}$ it the 1D PPP of the BSs in vertical street $k$.
      \item $x_{jk}$ is the distance from BS $j \in \Psi_{b,k}$ to the crossroad of street $k$ with the user street. 
  \end{itemize}
On the basis of these notations, the CF of the interference associated to diffraction is given by \eqref{proof2}.  Regarding the successive steps, (1) is obtained by computing the expectation of the fading coefficients, and by definition of $\Phi_F(\cdot,\cdot)$. (2) and (3) come from the definition of the PGFL \cite{baccelibook}, applied on $\Psi_{b,k}$ then on $\Psi_{VS}$.

  \section{Proof of Equation \eqref{eqCovprob}: coverage probability of a typical UE}
  \label{eqCovprobProof}
  Conditioned on the serving distance $r$ and on $(p)$, the propagation condition of the serving BS (LOS or NLOS), the coverage probability can be expressed as \eqref{proof3} where (1) is obtained by definition of the coverage probability and by rearranging terms and (2) is computed using the Gil-Pelaez theorem \cite{gil1951note}. The desired expression is finally given by \eqref{proof4} where (1) is obtained by averaging over the distibution of the serving distance and by using the total law of probability; (2) is obtained by developing the expressions of the conditioned probabilities.

\section*{Acknowledgment}
This work was supported by F.R.S.-FNRS under the EOS program (EOS project 30452698). Computational resources have been provided by the supercomputing facilities of the Université catholique de Louvain (CISM/UCL) and the Consortium des Équipements de Calcul Intensif en Fédération Wallonie Bruxelles (CÉCI) funded by the Fond de la Recherche Scientifique de Belgique (F.R.S.-FNRS) under convention 2.5020.11 and by the Walloon Region.

\ifCLASSOPTIONcaptionsoff
  \newpage
\fi

\bibliographystyle{IEEEtran}
\small
\bibliography{References}


\end{document}